\documentclass[12pt,prd]{revtex4}

\usepackage{amsmath}
\usepackage{graphicx}
\usepackage{amssymb}

\newcommand{\id}{{\bf 1}}
\newcommand{\arxiv}[1]{{arXiv:#1}}
\newcommand{\be}{\begin{equation}}
\newcommand{\ee}{\end{equation}}
\newcommand{\ba}{\begin{align}}
\newcommand{\ea}{\end{align}}
\newcommand{\nn}{\nonumber}

\newcommand{\chibar}{{\bar{\chi}}}

\newcommand{\Fint}{{\cal D}}

\newcommand{\Mhat}{\hat{M}}
\newcommand{\Tr}{\mathrm{Tr}}

\usepackage[normalem]{ulem}  % \sout{old text} for strikeout
\usepackage[dvips]{color} % For blue in-text comments and additions
\begin{document}

\hfill\parbox{3cm}
{
  KUNS-2402\\
  RIKEN-MP-49\\
  YITP-12-46
}

\title{
Strong-coupling Analysis of Parity Phase Structure \\
in Staggered-Wilson Fermions 
}

\author{Tatsuhiro Misumi}
\email{tmisumi@bnl.gov}
\affiliation{Physics Department, Brookhaven National Laboratory,
         Upton, New York 11973, USA}

\author{Takashi Z. Nakano}
\email{tnakano@yukawa.kyoto-u.ac.jp}
\affiliation{Department of Physics, Yukawa Institute for Theoretical Physics, Kyoto University,
         Kyoto 606-8502, Japan}

\author{Taro Kimura}
\email{tkimura@ribf.riken.jp}
\affiliation{Mathematical Physics Laboratory, RIKEN Nishina Center, 
         Saitama 351-0198, Japan}

\author{Akira Ohnishi}
\email{ohnishi@yukawa.kyoto-u.ac.jp}
\affiliation{Yukawa Institute for Theoretical Physics, Kyoto University,
         Kyoto 606-8502, Japan}

\begin{abstract}
We study strong-coupling lattice QCD with staggered-Wilson 
fermions, with emphasis on discrete symmetries and possibility of their 
spontaneous breaking. We perform hopping parameter 
expansion and effective potential analyses in the strong-coupling limit. 
From gap equations we find nonzero pion condensate in some range 
of a mass parameter, which indicates existence of the parity-broken 
phase in lattice QCD with staggered-Wilson fermions. We also find 
massless pions and PCAC relations around second-order phase boundary.
These results suggest that we can take a chiral limit by 
tuning a mass parameter in lattice QCD with staggered-Wilson 
fermions as with the Wilson fermion. 
\end{abstract}

\maketitle

\newpage

%%%%%%%%%%   Introduction   %%%%%%%%%%

\section{Introduction}
\label{sec:Intro}

Since the dawn of lattice field theory \cite{Wil}, the doubling problem
of fermions has been a notorious obstacle to lattice simulations. 
Among several prescriptions for this problem, the Wilson 
fermion simply bypasses the no-go theorem \cite{NN} by introducing 
a species-splitting mass term into the naive lattice fermion. This Wilson 
term is regarded as one example of  
``flavored-mass terms" which split original 16 fermion species into 
plural branches \cite{CKM1,CKM12}. It has been recently shown that the 
flavored-mass terms can also be constructed for staggered 
fermions \cite{KS, Suss, Sha} in Ref.~\cite{Adams1, Adams2, Hoel}. 
The original purpose of introducing these terms was establishment 
of the index theorem with staggered fermions \cite{Adams1}. 
A bonus here is that staggered fermions with 
the flavored-mass terms can be applied to lattice QCD simulations 
as Wilson fermion and an overlap kernel. One possible advantage 
of these novel formulations, called staggered-Wilson and staggered-overlap, 
is reduction of matrix sizes in the Dirac operators, 
which would lead to reduction of numerical costs in lattice simulations. 
The numerical advantage in the staggered-overlap fermion 
have been shown in \cite{PdF}. Now further study is required 
toward lattice QCD with these lattice fermions.

The purpose of this work is to reveal properties of staggered-Wilson 
fermions in terms of the parity-phase structure 
(Aoki phase) \cite{AokiP, AokiU1, CreutzW, SS, ACV, Sharpe}. 
As is well-known, the existence of the Aoki phase and the second-order 
phase boundary in Wilson-type fermions enables us to perform 
lattice QCD simulations by taking a chiral limit since the
critical behavior near the phase boundary reproduces massless 
pions and the PCAC relation. Besides, understanding the phase 
structure also gives practical information for the application of the 
overlap \cite{GW,Neu} and domain-wall \cite{Kap, FuSh} versions, 
both built on the Wilson-type kernel. Thus, in order to judge 
applicability of these new lattice fermions, it is essential to investigate 
the Aoki phase in the staggered-Wilson fermions. The phase structure for the 
staggered-Wilson fermion was first studied by using the Gross-Neveu 
model in Ref.~\cite{CKM2, MCKNO} and the present paper shows further 
investigation of this topic.

In this paper, we investigate strong-coupling lattice QCD \cite{KaS} 
with emphasis on parity-phase structure for two types 
of staggered-Wilson fermions \cite{Adams2, Hoel}. Firstly we discuss discrete
symmetries of staggered-Wilson fermions, and show that physical parity and
charge conjugation can be defined in both cases while hypercubic symmetry
depends on types of staggered-Wilson fermions. 
Secondly, we perform hopping-parameter expansion and 
effective potential analysis for meson fields in the strong-coupling limit.
For this purpose, we develop a method to derive the effective potential 
for lattice fermions with multiple-hopping terms. The gap equations show 
that pion condensate becomes non-zero in some range of a mass parameter, 
which indicates that parity-broken phase appears in this range. We 
also study meson masses around the second-order phase boundary, 
and find that massless pions and PCAC relations are reproduced. 
Lastly, we discuss parity-flavor symmetry breaking for
2-flavor cases. These results suggest that we can take a chiral limit by 
tuning a mass parameter in lattice QCD with staggered-Wilson 
fermions as with the Wilson fermion.

This paper is organized as follows.
In Sec.~\ref{sec:SWF}, we review staggered flavored-mass terms and 
two types of staggered-Wilson fermions. 
In Sec.~\ref{sec:Sym}, we study discrete symmetries of 
staggered-Wilson fermions.
In Sec.~\ref{sec:HPE}, we study hopping parameter expansion
in lattice QCD with these fermions.  
In Sec.~\ref{sec:EPA}, we investigate Aoki phase structure by
effective potential analysis. 
In Sec.~\ref{sec:Tf}, we discuss parity-flavor symmetry breaking in 
two-flavor cases.
In Sec.~\ref{sec:SD}, we devote ourselves to a summary 
and discussion.

%%%%%%%%%%%%%% Staggered Wilson fermions %%%%%%%%%%

\section{Staggered-Wilson fermions}
\label{sec:SWF}

Before looking into staggered-Wilson fermions, we review the Wilson fermion
and its relatives. The Wilson term splits 16 species of naive fermions 
into 5 branches with 1, 4, 6, 4 and 1 fermions. We call this kind of 
species-splitting terms ``flavored-mass terms". As shown in \cite{CKM1}, 
there are 4 types of flavored-mass terms for naive fermion which satisfy 
$\gamma_{5}$ hermiticity. ($\gamma_{5}$ in the naive 
fermion is flavored such as 
$\gamma_{5}\otimes(\tau_{3}\otimes\tau_{3}\otimes\tau_{3}\otimes\tau_{3})$ 
in the spin-flavor representation.)  All these terms with proper mass shifts 
lead to a second derivative term as $\sim a\int dx^{4} \bar{\psi}D^{2}_{\mu}\psi$ 
up to $\mathcal{O}(a^2)$ errors. 
Thus we can regard them as cousins of Wilson fermion. 

There are also non-trivial flavored-mass terms for staggered fermions, 
which split 4 tastes into branches and satisfy $\gamma_{5}$ hermiticity. 
Since $\gamma_{5}$ is expressed in spin-taste representation as 
$\gamma_{5}\otimes\gamma_{5}$ in this case, we only have two 
flavored-mass terms satisfying $\gamma_{5}$ hermiticity: $\id\otimes\gamma_{5}$ 
and $\id\otimes \sigma_{\mu\nu}$. (For larger discrete symmetry one needs to
take a proper sum for $\mu,\nu$ in the latter case.) These spin-flavor 
representations translate into four- and two-hopping terms in
the one-component staggered action up to $\mathcal{O}(a)$ errors.
The first type is given by
\begin{equation}  
M_A= \epsilon\sum_{sym} \eta_{1}\eta_{2}\eta_{3}\eta_{4}
C_{1}C_{2}C_{3}C_{4}
= (\id \otimes \gamma_{5}) + \mathcal{O}(a)
\ ,
\label{AdamsM}
\end{equation}
with
\begin{align}
(\epsilon)_{xy}&=(-1)^{x_{1}+...+x_{4}}\delta_{x,y}
\ ,
\\
(\eta_{\mu})_{xy}&=(-1)^{x_{1}+...+x_{\mu-1}}\delta_{x,y}
\ ,
\\
C_{\mu}&=(V_{\mu}+V_{\mu}^{\dag})/2
\ ,  
\\
(V_{\mu})_{xy}&=U_{\mu,x}\delta_{y,x+\mu}
\ .
\end{align}
The second type is given by
\begin{align}  
M_H&={i\over{\sqrt{3}}}(\eta_{12}C_{12}+\eta_{34}C_{34}+\eta_{13}C_{13}+\eta_{42}C_{42}
+\eta_{14}C_{14}+\eta_{23}C_{23})
\ ,
\nonumber\\ 
&= [\id \otimes (\sigma_{12}+\sigma_{34}+\sigma_{13}+\sigma_{42}+\sigma_{14}+\sigma_{23}) ] + \mathcal{O}(a)
\ ,
\label{HoelM}
\end{align}
with
\begin{align}
(\eta_{\mu\nu})_{xy}&=\epsilon_{\mu\nu}\eta_{\mu}\eta_{\nu}\delta_{x,y}
\ ,
\\
(\epsilon_{\mu\nu})_{xy}&=(-1)^{x_{\mu}+x_{\nu}}\delta_{x,y}
\ , 
\\
C_{\mu\nu}&=(C_{\mu}C_{\nu}+C_{\nu}C_{\mu})/2
\ .
\end{align}
We refer to $M_A$ and $M_H$ as the Adams- \cite{Adams1} and 
Hoelbling-type \cite{Hoel}, respectively. The former splits the 4 tastes 
into two branches with positive mass and the other two with negative mass.
These two branches correspond to $+1$ and $-1$ eigenvalues of 
$\gamma_{5}$ in the taste space.
The latter splits them into one with positive mass, two with zero mass
and the other one with negative mass. 
We note that $M_{A}$ and $M_{H}$ are also derived from the flavored
mass terms for naive fermions through spin-diagonalization as shown in \cite{CKM1}.   
Now we introduce a Wilson parameter $r=r\delta_{x,y}$ and shift  mass 
as in Wilson fermions \cite{Hoel}.
Then the Adams-type staggered-Wilson fermion action is given by
\begin{align}  
S_{\rm A}\,&=\,\sum_{xy}\bar{\chi}_{x}[\eta_{\mu}D_{\mu}
+r(1+M_A)+M]_{xy}\chi_{y}
\ ,
\label{AdamsS}
\\
D_{\mu}&={1\over{2}}(V_{\mu}-V^\dagger_{\mu}) 
\ .
\end{align}
Here $M$ stands for the usual taste-singlet mass ($M=M\delta_{x,y}$). 
The Hoelbling-type fermion action is given by
\begin{align}  
S_{\rm H}\,=\,\sum_{xy}\bar{\chi}_{x}[\eta_{\mu}D_{\mu}
+r(2+M_H)+M]_{xy}\chi_{y}
\ .
\label{HoelS}
\end{align}
It is obvious that these lattice fermions have possibility to be two- 
or one-flavor Wilson fermions.
In lattice QCD simulation with these fermions, the mass parameter $M$ will be
tuned to take a chiral limit as Wilson fermions. For some negative value of the 
mass parameter: $-1<M<0$ for Adams-type and $-2<M<0$ for Hoelbling-type, 
we obtain two-flavor and one-flavor overlap fermions respectively by 
using the overlap formula in principle.

%%%%%%%%%%%%% Symmetry %%%%%%%%%%%

\section{Discrete Symmetries}
\label{sec:Sym}

In this section, we discuss discrete symmetry of staggered-Wilson fermions.
A potential problem with staggered-Wilson fermions in lattice QCD 
is discrete symmetry breaking. As discussed in \cite{Adams2, Hoel}, 
the discrete symmetry possessed by the original staggered fermion 
is broken to its subgroup both in the Adams-type and Hoelbling-type actions.
We below list all the staggered discrete symmetries (shift, axis reversal, rotation 
and charge conjugation), and look into their status in staggered-Wilson fermions. 
Shift transformation is given by
\begin{align}
\mathcal{S}_{\mu}:\,\,\chi_{x} \to \zeta_{\mu}(x)\chi_{x+\hat{\mu}}, \,\,\,\,
\bar{\chi}_{x} \to \zeta_{\mu}(x)\bar{\chi}_{x+\hat{\mu}},\,\,\,\,
U_{\nu,x} \to U_{\nu, x+\hat{\mu}}
\ ,
\label{shift1}
\end{align}
with $\zeta_{1}(x)=(-1)^{x_{2}+x_{3}+x_{4}}$, $\zeta_{2}(x)=(-1)^{x_{3}+x_{4}}$,
$\zeta_{3}(x)=(-1)^{x_{4}}$ and $\zeta_{4}(x)=1$.
It is obvious that this transformation flips the sign of both flavored-mass terms.
The Adams type is invariant under the two-shift subgroup as $x \to x+\hat{1}\pm\hat{\mu}$ 
while the Hoelbling type is invariant under four-shift subgroup as 
$x\to x+\hat{1}\pm\hat{2}\pm\hat{3}\pm\hat{4}$.
Note that these subgroups include the doubled shift $x\to x+2\hat{\mu}$ as their subgroup. 
The axis reversal transformation is given by,
\begin{align}
\mathcal{I}_{\mu}:\,\,\chi_{x} \to (-1)^{x_{\mu}}\chi_{Ix}, \,\,\,\,
\bar{\chi}_{x} \to (-1)^{x_{\mu}}\bar{\chi}_{Ix},\,\,\,\,
U_{\nu,x} \to U_{\nu, Ix}
\ ,
\label{axis1}
\end{align}
with $I=I^{\mu}$ is the axis reversal $x_{\mu}\to -x_{\mu}$, $x_{\rho}\to x_{\rho}$,
$\rho\not= \mu$. It again flips the sign of the flavored-mass terms.
The staggered rotational transformation is given by
\begin{align}
\mathcal{R}_{\rho\sigma}:\,\,\chi_{x} \to S_{R}(R^{-1}x)\chi_{R^{-1}x},\,\,\,\,
\bar{\chi_{x}} \to S_{R}(R^{-1}x)\bar{\chi}_{R^{-1}x},\,\,\,\,
U_{\nu, x} \to U_{\nu, Rx}
\ ,
\label{rot1}
\end{align}
where $R_{\rho\sigma}$ is the rotation $x_{\rho}\to x_{\sigma}$, $x_{\sigma}\to -x_{\rho}$,
$x_{\tau}\to x_{\tau}$, $\tau \not= \rho, \sigma$ and 
$S_{R}(x)={1\over{2}}[1\pm\eta_{\rho}(x)\eta_{\sigma}(x)\mp\zeta_{\rho}(x)\zeta_{\sigma}(x)
+\eta_{\rho}(x)\eta_{\sigma}(x)\zeta_{\rho}(x)\zeta_{\sigma}(x)]$ with 
$\rho$\hspace{0.3em}\raisebox{0.4ex}{$<$}\hspace{-0.75em
}\raisebox{-0.7ex}{$>$}\hspace{0.3em}$\sigma$.
It is notable that the Adams-type fermion keeps this staggered rotation symmetry 
while the Hoelbling type loses it.
The staggered charge conjugation transformation is given by
\begin{equation}
\mathcal{C}:\,\,\chi_{x}\to\epsilon_{x}\bar{\chi}_{x}^{T},\,\,\,\,
\bar{\chi}_{x}\to-\epsilon_{x}\chi_{x}^{T},\,\,\,\,
U_{\nu,x} \to U_{\nu,x}^{*}
\ .
\label{C0}
\end{equation}
The Adams-type fermion keeps this symmetry while the Hoelbling type loses it.

We next elucidate residual subgroups possessed by 
staggered-Wilson fermions, and discuss how to define
physical discrete symmetries as Parity, Charge conjugation and Hypercubic
symmetry. For this purpose we separate spin and flavor rotations
in the above transformations. Here we utilize the momentum-space representation
in \cite{GS,DS}. In this representation we can define two set of clifford generators 
$\Gamma_{\mu}$ and $\Xi_{\mu}$, which operate on spinor and flavor 
spaces in the momentum-space field $\phi(p)$, respectively.
(Details are shown in Appendix.~\ref{SFS}.)
Then the shift transformation translates into 
\begin{equation}
\mathcal{S}_{\mu}:\,\,\phi(p)\,\,\to\,\, \exp(ip_{\mu})\Xi_{\mu}\,\phi(p)
\ .
\label{shift2}
\end{equation}
The axis reversal translates into
\begin{equation}   
\mathcal{I}_{\mu}:\phi(p)\,\,\to\,\,\Gamma_{\mu}\Gamma_{5}\Xi_{\mu}\Xi_{5}\,\phi(Ip)
\ .
\label{axis2}
\end{equation}
The rotational transformation translates into
\begin{equation}
\mathcal{R}_{\rho\sigma}:\,\,\phi(p)\,\,\to\,\,\exp({\pi\over{4}}\Gamma_{\rho}\Gamma_{\sigma})
\exp({\pi\over{4}}\Xi_{\rho}\Xi_{\sigma})\,\phi(R^{-1}p)
\ .
\label{rot2}
\end{equation}
By using this representation we figure out 
residual discrete symmetries of staggered-Wilson fermions as follows.
We first consider parity.
Both staggered-Wilson fermions are invariant under
\begin{equation} 
\mathcal{I}_{s}\mathcal{S}_{4}\sim
\exp(ip_{4})\Gamma_{1}\Gamma_{2}\Gamma_{3}\Gamma_{5}\,\phi(-{\bf p},p_{4})\sim
\exp(ip_{4})\Gamma_{4}\,\phi(-{\bf p},p_{4})
\ ,  
\label{parity}
\end{equation}
with $\mathcal{I}_{s}\equiv \mathcal{I}_{1}\mathcal{I}_{2}\mathcal{I}_{3}$.
This is essentially the continuum parity transformation \cite{GS}. 
In the continuum limit the phase factor disappears and it results in the
continuum parity transformations $P : \psi(p)\to\gamma_{4}\psi(-{\bf p},p_{4})$ 
for the Dirac fermion. 
We thus conclude both staggered-Wilson fermions possess symmetry leading to
physical parity symmetry $P$. We note the simple combination of $\mu$-shift 
and $\mu$-axis reversal (shifted-axis reversal:$\mathcal{S}_{\mu}\mathcal{I}_{\mu}$)
is also a symmetry of both fermions.

We next consider physical charge conjugation.
In the case of Adams fermion the staggered charge conjugation symmetry
$\mathcal{C}$ in Eq.~(\ref{C0}) remains intact. Thus, physical charge conjugation
for the two-flavor branch can be formed in a similar way to usual staggered 
fermions as shown in \cite{GS} ($C\sim\mathcal{C}\mathcal{S}_{2}\mathcal{S}_{4}
\mathcal{I}_{2}\mathcal{I}_{4}$).
On the other hand, the Hoelbling type breaks $\mathcal{C}$.
In this case, however, we can define another charge conjugation 
by combining $\mathcal{C}$ with rotation transformation as
\begin{equation}
\mathcal{C}_{T} : \,\,\mathcal{R}_{21}\mathcal{R}_{13}^{2}\mathcal{C}
\ .
\label{RRRC}
\end{equation}
The Hoelbling action is invariant under this transformation.
By using this symmetry we can define physical charge conjugation $C$
for one-flavor branch as in the Adams type. 
We thus conclude that both staggered-Wilson fermions have proper 
charge conjugation symmetry.

%%%%%%%%%%%%%%%%%%%%%%%%%%%%%%%%%%%%%%%%%%%%%%%%%%%%%%%%%%%%%%%%%%%%%%%%%%%%%%%%
\begin{table*}[hbt]
\caption{Symmetries of staggered-Wilson fermions.}
\label{Table:sym}
\begin{tabular}{c|ccccccc}
\hline
\hline
& $N_f$ & \,\,\,\,\,$\mathcal{S}$$\&$$\mathcal{I}$-subgroup & \,\,\,\,\,$\mathcal{R}$-subgroup 
& $P$ &  $C$ & $SW_{4}$ \\ \hline
Staggered & $4$ & $\mathcal{S}_{\mu}$,\,\,\,\,$\mathcal{I}_{\mu}$ 
& $\mathcal{R}_{\mu\nu}$ & $\bigcirc$ & $\bigcirc$ & $\bigcirc$ \\    
Adams & $2$ & $\mathcal{S}_{\mu}\mathcal{S}_{\nu}$,\,
$\mathcal{S}_{\mu}\mathcal{I}_{\mu}$ & $\mathcal{R}_{\mu\nu}$ 
& $\bigcirc$ & $\bigcirc$ & $\bigcirc$ \\ 
Hoelbling & $1$ & $\mathcal{S}_{\mu}\mathcal{S}_{\nu}\mathcal{S}_{\rho}\mathcal{S}_{\sigma}$,
\,$\mathcal{S}_{\mu}\mathcal{I}_{\mu}$ & $\mathcal{R}_{\mu\nu}\mathcal{R}_{\rho\sigma}$ 
& $\bigcirc$ & $\bigcirc$ & $\times$ \\   
\hline
\hline
\end{tabular}
\end{table*}
%%%%%%%%%%%%%%%%%%%%%%%%%%%%%%%%%%%%%%%%%%%%%%%%%%%%%%%%%%%%%%%%%%%%%%%%%%%%%%%%

We lastly consider hypercubic symmetry.
In staggered fermions, the rotation Eq. (\ref{rot1}) and axis reversal Eq. (\ref{axis1})
form hypercubic symmetry \cite{KilSha}, which enhances to Euclidian Lorentz symmetry
in the continuum limit. In the case of the Adams-type fermion, the action is invariant 
under the rotation Eq. (\ref{rot1}) and the shifted-axis reversal 
$\mathcal{S}_{\mu}\mathcal{I}_{\mu}$. These two symmetries can form 
proper hypercubic symmetry $SW_{4}$ in this case. Thus we conclude that Adams fermion 
recovers Lorentz symmetry in the continuum.
In the Hoelbling-type formulation, the action breaks the rotation symmetry 
into its subgroup called doubled rotation  \cite{Hoel}, which is given by
\begin{equation}
\mathcal{R}_{\rho\sigma}\mathcal{R}_{\mu\nu}\sim
\exp[{\pi\over{4}}(\Gamma_{\rho}\Gamma_{\sigma}+\Gamma_{\mu}\Gamma_{\nu})]
\exp[{\pi\over{4}}(\Xi_{\rho}\Xi_{\sigma}+\Xi_{\mu}\Xi_{\nu})]
\phi(R_{\rho\sigma}^{-1}R_{\mu\nu}^{-1}p)
\ ,
\label{DR}
\end{equation}
where $(\mu,\nu,\sigma,\rho)$ is any permutation of ($1,2,3,4$).
It is also invariant under sequential transformations of
($\mu,\nu$ rotation), ($\nu,\mu$ rotation), ($\mu$ shift) and ($\nu$ shift) as
\begin{equation}
\mathcal{S}_{\nu}\mathcal{S}_{\mu}\mathcal{R}_{\nu\mu}\mathcal{R}_{\mu\nu}\sim
\exp(ip_{\mu}+ip_{\nu})\Gamma_{\mu}\Gamma_{\nu}\,\phi(\tilde{p})
\ ,
\label{SDR}
\end{equation}
with $\tilde{p}_{\mu,\nu}=-p_{\mu,\nu}, \tilde{p}_{\tau}=p_{\tau}$, $\tau\not=\mu,\nu$. 
The loss of rotation symmetry indicates that we cannot define hypercubic symmetry
in the Hoelbling fermion. It implies that it could not lead to a correct continuum theory,
and we would need to tune parameters to restore Lorentz symmetry.
Indeed the recent study on symmetries of staggered-Wilson fermions by Sharpe \cite{Steve}
reports that recovery of Lorentz symmetry requires fine-tuning of coefficients 
in the gluonic sector in lattice QCD with Hoelbling fermion.

To summarize, Adams fermion possesses physical parity, charge conjugation
and hypercubic symmetries while Hoelbling fermion loses hypercubic as shown 
in Table.~\ref{Table:sym}. 
It seems that Hoelbling fermion cannot be straightforwardly applied to
lattice QCD while Adams type can. 
We note that both staggered-Wilson fermions have proper parity symmetry,
and we can discuss spontaneous parity symmetry breaking.
Moreover, we may find some symptom due to Lorentz symmetry breaking 
in Hoelbling fermion in the parity-phase structure.
This is another motivation to study parity-phase structure in 
strong-coupling lattice QCD. 

In the end of this section, we comment on special symmetries in 
staggered-Wilson fermions without mass shift. Hoelbling fermion
without the mass shift ($\eta_{\mu}D_{\mu}+M_{H}$) possesses 
special charge conjugation symmetry ($\mathcal{C}_{T}': \chi\to\bar\chi,\,
\bar{\chi}\to\chi$). This topic is beyond the scope of this work, but we note
that it can do good to two flavors in the central branch. Use of the
central branch is intensively discussed in \cite{Rev, PdF}.

%%%%%%%%%%% Hopping Parameter Expansion %%%%%%%%%%

\section{Hopping Parameter Expansion}
\label{sec:HPE}
In this section we investigate parity-phase structure in lattice QCD with 
staggered-Wilson fermions in the framework of hopping parameter 
expansion (HPE) in the strong-coupling regime \cite{AokiP}. 
In the hopping parameter expansion, we treat a mass term as a leading action
while we perturbatively treat hopping terms.
We thus perform expansion by a hopping parameter which essentially corresponds 
to inverse of a mass parameter.
By using self-consistent equations we derive one- and two-point 
functions, and calculate meson condensates 
and meson mass for two types of staggered-Wilson fermions.
We for simplicity drop the flavor indices until we discuss the 
two-flavor case in details in Sec.~\ref{sec:Tf}. However it is easy to recover 
the flavor indices for the field $\chi_{f}$, 
the mass parameter $M_{f}$ and  the condensate $\Sigma_{f}$ ($f=1,2,...$).

\subsection{Hoelbling type}
\label{subsec:HPE-H}

%%%%%%%%
\begin{figure}
 \begin{center}
  \includegraphics[height=6cm]{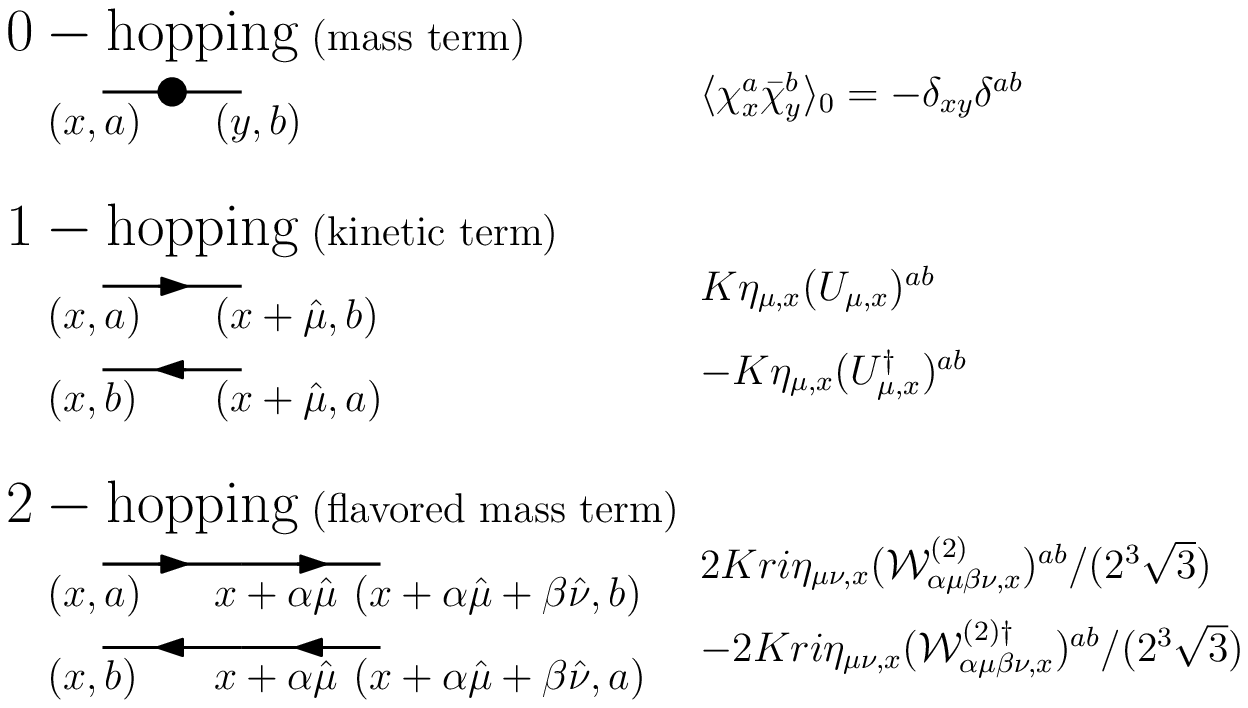}\,\,\,\,
   \end{center}
 \caption{Feynman rules in hopping parameter expansion (HPE) with the Hoelbling-type staggered-Wilson fermion. $a$ and $b$ stand for the color indices. $\mathcal{W}_{\alpha\mu\beta\nu,x}^{(2)}$ is given in Table \ref{Table:U2}.}
 \label{FR-H}
\end{figure}
%%%%%%%%%
%%%%%%%%%%%%%%%%%%%%%%%%%%%%%%%%%%%%%%%%%%%%%%%%%%%%%%%%%%%%%%%%%%%%%%%%%%%%%%%%
\begin{table*}[hbt]
\caption{Concrete forms of $\mathcal{W}_{\alpha\mu\beta\nu,x}^{(2)}$ in Fig. \ref{FR-H}.
}
\label{Table:U2}
\begin{tabular}{ccc}
\hline
\hline
$\alpha$ & $\beta$ & $\mathcal{W}_{\alpha\mu\beta\nu,x}^{(2)}$ \\ \hline
$+$ & $+$ & $U_{\mu,x}U_{\nu,x+\hat{\mu}}$ \\ 
$-$ & $-$ & $U_{\mu,x-\hat{\mu}}^\dagger U_{\nu,x-\hat{\mu}-\hat{\nu}}^\dagger$ \\ 
$+$ & $-$ & $U_{\mu,x}U_{\nu,x+\hat{\mu}-\hat{\nu}}^\dagger$ \\ 
$-$ & $+$ & $U_{\mu,x-\hat{\mu}}^\dagger U_{\nu,x-\hat{\mu}}$ 
\\ 
\hline
\hline
\end{tabular}
\end{table*}
%%%%%%%%%%%%%%%%%%%%%%%%%%%%%%%%%%%%%%%%%%%%%%%%%%%%%%%%%%%%%%%%%%%%%%%%%%%%%%%%
%%%%%%%%
\begin{figure}
 \begin{center}
  \includegraphics[height=6cm]{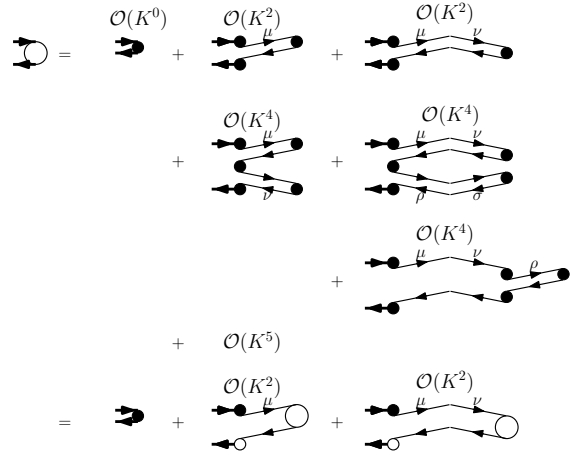}
 \end{center}
 \caption{Feynman diagram for mesonic one-point functions
 in the $\mathcal{O}(K^{3})$ self-consistent equation of HPE with the Hoelbling fermion. 
 Black circles stand for the leading one-point function $\langle \chi_x \bar{\chi}_x \rangle_0$ 
 while white circles stand for $\langle \chi_x \bar{\chi}_x \rangle$ 
 which include next-leading and higher hopping terms. 
 By summing up higher contributions, we obtain the second equality.}
 \label{One-H}
\end{figure}
%%%%%%%%%

We begin with the Hoelbling-type fermion, which contains two-hopping terms
in the action. One reason that we start with Hoelbling type regardless of the lower
rotation symmetry is that 2-hopping calculation can be a good exercise for
4-hopping case in Adams type.
To perform the HPE for the Hoelbling-type fermion, 
we rewrite the action Eq.~(\ref{HoelS}) by redefining $\chi \rightarrow \sqrt{2K} \chi$ 
with the hopping parameter $K=1/[2(M+2r)]$,
\begin{equation}
S =
\sum_{x} \chibar_x \chi_x + 2K \sum_{x, y} \chibar_{x}(\eta_{\mu}  
D_\mu )_{xy}\chi_{y}+ 2 K r \sum_{x,y} \chibar_x (M_{H})_{xy} \chi_y
\ ,
\label{HPE-Hoel}
\end{equation}
where $M_{H}$ is given by Eq.~(\ref{HoelM}).
The plaquette action is $1/g^2$ term and we can omit it in the strong-coupling limit.
In this section, we derive one- and two-point functions by using 
a $\mathcal{O}(K^{3})$ self-consistent equation:
Solving this equation leads to truncation of diagrams as the ladder approximation
having all diagrams to $\mathcal{O}(K^3)$ are taken into account.
More precisely, this approximation does not take account of all diagrams,
but it successfully includes certain kinds of diagrams to all orders of $K$
thanks to a self-consistent approach. 
We thus expect that it works to figure out existence of Aoki phase.
We note that this approximation especially works well for a small hopping 
parameter $K\ll1$. 
In Fig.~\ref{FR-H}, we depict Feynman rules in the HPE for this fermion.
The fundamental Feynman rules contain contributions from 
0-hopping (mass term), 1-hopping (kinetic term) and 2-hopping (flavored-mass term) terms.

First, by using these Feynman rules, we derive meson condensates from an one-point
function of the meson operator $\mathcal{M}_{x}=\bar{\chi}_{x}\chi_{x}$ 
in the mean-field approximation.
The one-point function is defined as
\begin{align}
\langle \chi_x^a \bar{\chi}_x^b \rangle
\equiv& - \delta_{ab} \Sigma_x
= \frac{\int \mathcal{D}[\chi,\bar{\chi},U] \chi_x^a \bar{\chi}_x^b\ e^{S}
}{
\int \mathcal{D}[\chi,\bar{\chi},U] e^{S}
}\ .
\end{align}
Note that we use the partition function $Z=\int \mathcal{D}[\chi,\bar{\chi},U] e^{S}$, 
not $Z=\int \mathcal{D}[\chi,\bar{\chi},U] e^{-S}$,
following the convention for the partition function in the strong-coupling analysis \cite{KaS}. 
The leading term in the hopping parameter expansion is given by
\begin{align}
\langle \chi_x^a \bar{\chi}_x^b \rangle_0
= \frac{\int \mathcal{D}[\chi,\bar{\chi},U]\, \chi_x^a \bar{\chi}_x^b\ e^{S_0}}
       {\int \mathcal{D}[\chi,\bar{\chi},U]\, e^{S_0}}
= - \delta^{ab}
\ ,
\end{align}
where $S_0=\sum_x \bar{\chi}_x \chi_x$.
By using the Feynman rules, we can evaluate the diagrams in Fig.~\ref{One-H}.
\begin{align}
\langle \chi_x^a \bar{\chi}_x^b \rangle 
& \equiv - \delta^{ab} \Sigma_x  
\nn \\  
&= \langle \chi_x^a \bar{\chi}_x^b \rangle_0 
\nn \\  
&
 + \sum_{\pm \mu} 
   (-1) (K \eta_{\mu,x})^2 
   \langle (\chi^a \bar{\chi})_x \rangle_0 
   U_{\mu,x} 
   \langle (\chi \bar{\chi})_{x+\hat{\mu}} \rangle_0 
   U_{\mu,x}^\dagger
   \langle (\chi \bar{\chi}^b)_x \rangle_0 
\nn \\  
&
 + \sum_{\substack{\pm \mu,\pm \nu \\ (\mu \neq \nu)}} 
   (-1) \left( 2K r i \eta_{\mu \nu,x} \displaystyle \frac {1}{2^3 \sqrt{3}} \right)^2 
   \langle (\chi^a \bar{\chi})_x \rangle_0 
   \mathcal{W}_{\mu\nu,x}^{(2)} 
   \langle (\chi \bar{\chi})_{x+\hat{\mu}+\hat{\nu}} \rangle_0 
   \mathcal{W}_{\mu\nu,x}^{(2)\dagger} 
   \langle (\chi \bar{\chi}^b)_x \rangle_0 
\nn \\  
&
 + \sum_{\pm \mu,\pm \nu} 
  (-1) (K \eta_{\mu,x})^2 (-1) (K \eta_{\nu,x})^2
  \langle (\chi^a \bar{\chi})_x \rangle_0 
  U_{\mu,x} 
  \langle (\chi \bar{\chi})_{x+\hat{\mu}} \rangle_0 
  U_{\mu,x}^\dagger
  \langle (\chi \bar{\chi})_x \rangle_0 U_{\nu,x} 
\nn \\  
& \times
  \langle (\chi \bar{\chi})_{x+\hat{\nu}} \rangle_0 
  U_{\nu,x}^\dagger
  \langle (\chi \bar{\chi}^b)_x \rangle_0 
\nn \\  
&
 + \sum_{\substack{\pm \mu,\pm \nu ,\pm \rho, \pm \sigma \\ (\mu \neq \nu, \rho \neq \sigma)}} 
   (-1) \left( 2K r i \eta_{\mu \nu,x} \displaystyle \frac {1}{2^3 \sqrt{3}} \right)^2 
   (-1) \left( 2K r i \eta_{\rho \sigma,x} \displaystyle \frac {1}{2^3 \sqrt{3}} \right)^2 
\nn \\  
& \times
   \langle (\chi^a \bar{\chi})_x \rangle_0 
   \mathcal{W}_{\mu\nu,x}^{(2)} 
   \langle (\chi \bar{\chi})_{x+\hat{\mu}+\hat{\nu}} \rangle_0 
   \mathcal{W}_{\mu\nu,x}^{(2)\dagger} 
   \langle (\chi \bar{\chi})_x \rangle_0  
   \mathcal{W}_{\rho\sigma,x}^{(2)} 
   \langle (\chi \bar{\chi})_{x+\hat{\rho}+\hat{\sigma}} \rangle_0 
   \mathcal{W}_{\rho\sigma,x}^{(2)\dagger} 
   \langle (\chi \bar{\chi}^b)_x \rangle_0 
\nn \\  
&
 + \sum_{\substack{\pm \mu,\pm \nu ,\pm \rho \\ (\mu \neq \nu)}} 
   (-1) \left( 2K r i \eta_{\mu \nu,x} \displaystyle \frac {1}{2^3 \sqrt{3}} \right)^2 
   (-1) \left( K \eta_{\rho,x} \right)^2 
   \langle (\chi^a \bar{\chi})_x \rangle_0  
   \mathcal{W}_{\mu\nu,x}^{(2)} 
\nn \\  
& \times
   \langle (\chi \bar{\chi})_{x+\hat{\mu}+\hat{\nu}} \rangle_0 
   U_{\rho,x+\hat{\mu}+\hat{\nu}}
   \langle (\chi \bar{\chi})_{x+\hat{\mu}+\hat{\nu}+\hat{\rho}} \rangle_0 
   U_{\rho,x+\hat{\mu}+\hat{\nu}}^\dagger
   \langle (\chi \bar{\chi})_{x+\hat{\mu}+\hat{\nu}} \rangle_0 
   \mathcal{W}_{\mu\nu,x}^{(2)\dagger} 
   \langle (\chi \bar{\chi}^b)_x \rangle_0 
\nn \\  
& + \mathcal{O}(K^5)
\ ,
\label{Eq:HPE-Hoelbling1-detail}
\end{align}
where $(\chi \bar{\chi})_x$ stands for $\chi_x \bar{\chi}_x$ and $\mathcal{W}_{\mu\nu,x}^{(2)}=\mathcal{W}_{+\mu+\nu,x}^{(2)}$ 
in Table.~\ref{Table:U2}.
Note that we consider only connected diagrams in Fig.~\ref{One-H}. 
By summing higher hopping terms,
the one-point function is obtained as shown 
in Fig.~\ref{One-H}, which is given by
\begin{equation}
- \Sigma_x  \equiv - \langle \mathcal{M}_x \rangle =
 - \langle \mathcal{M}_x \rangle_0 
 + 2K^2 \sum_\mu \Sigma_x \Sigma_{x+\hat{\mu}} 
 - 2 \cdot \displaystyle \frac {1}{24} (Kr)^2 \sum_{\mu \neq \nu} \Sigma_x 
 \Sigma_{x+\hat{\mu}+\hat{\nu}}
\ .
\label{Eq:HPE-Hoelbling1}
\end{equation}
The equation contains terms to $\mathcal{O}(K^{2})$, and $\mathcal{O}(K^{3})$ diagrams 
are found to vanish due to cancellation between the diagrams. Here 
we solve it in a self-consistent way for condensate $\Sigma$ 
within mean-field approximation. We here assume 
$\Sigma_x=\sigma_x +i \epsilon_x \pi_x$ as
the condensate. 
$\sigma$ and $\pi$ correspond to chiral and pion condensates, respectively.
We substitute this form of $\Sigma_{x}$ in Eq.~(\ref{Eq:HPE-Hoelbling1}) 
and obtain a self-consistent equation 
\begin{equation}
- \left( \sigma +i \epsilon_x \pi \right)=-1 + 2K^2 \cdot 4 \left( \sigma^2 
+ \pi^2 \right) -2 \cdot \displaystyle \frac {1}{24} (Kr)^2 \cdot 4 \cdot 3 
\left( \sigma +i \epsilon_x \pi \right)^2
\ ,
\label{HPE-HoelSelf1}
\end{equation}
which yields $- \sigma = -1 + 16 K^2 \pi^2$ and 
$- i \pi = - 8K^2 \cdot 2 i \sigma \pi$. For simplicity, we have set $r=2\sqrt{2}$ 
to make the equation Eq. (\ref{HPE-HoelSelf1}) simpler. 
Of course we can also discuss for other values of $r$ in general.
 
Now we have two solutions depending on $\pi=0$ 
or $\pi\not=0$: For $\pi=0$, we have a trivial solution $\sigma=1$.
For $\pi \neq 0$, we have a solution as
\begin{equation}
\sigma = \displaystyle \frac{1}{16K^2},\,\,\,\,\,\,\,\,\,\,\,\,\,\,
\pi = \pm \sqrt{ \displaystyle \frac{1}{16K^2} \left( 1- \displaystyle 
\frac{1}{16K^2} \right) }
\ .
\label{cond}
\end{equation}
Nonzero pion condensate implies spontaneous parity breaking 
for the range $\mid{K}\mid > 1/4$. The sign of the pion condensate 
in Eq. (\ref{cond}) reflects the $Z_2$ parity symmetry of the theory.
Thus the parity-broken phase, if it exists, appears in a parameter range 
$-4\sqrt{2}-2<M<-4\sqrt{2}+2$ in the strong-coupling limit. We note 
that the critical hopping parameter $|K_c|=1/4$ is small, 
and we speculate that the $\mathcal{O}(K^{3})$ 
self-consistent equation is valid around the value.

\begin{figure}
 \begin{center}
  \includegraphics[height=4cm]{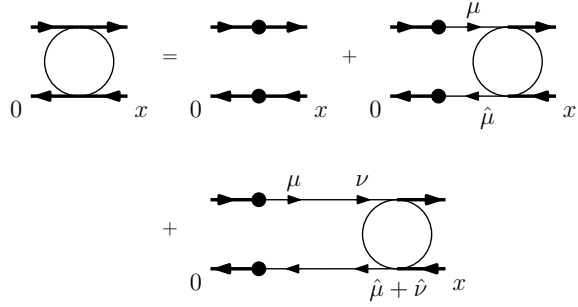}
 \end{center}
 \caption{Feynman diagram for mesonic two-point functions for
 $\mathcal{O}(K^{3})$ self-consistent equation with the Hoelbling fermion. }
 \label{Two-H}
\end{figure}

Next, we discuss the meson mass from a two-point function of the meson operator 
$\mathcal{S}(0,x)\equiv\langle\mathcal{M}_{0}\mathcal{M}_{x}\rangle$. 
From Fig.~\ref{Two-H}, we derive the following $\mathcal{O}(K^{3})$ equation 
for a two-point function.
%%%%%%%%
%%%%%%%%%
\begin{align}
\mathcal{S}(0,x) 
 = &\langle \chibar_0^a\chi_0^a \chibar_x^b \chi_x^b \rangle
\nn \\
 = &- \delta_{0x} N_c
\nn \\
 &-K^2 \langle \chibar_0^a \chi_0^a \chibar_0^c (\eta_{\mu,0})^2 
 \biggl[
   U_{\mu,0}^{cd} \chi_{\hat{\mu}}^d \chibar_{\hat{\mu}}^e 
   (U_{\mu,0}^{\dagger})^{ef}
 + (U_{\mu,-\hat{\mu}}^{\dagger})^{cd} 
   \chi_{-\hat{\mu}}^d 
 \chibar_{-\hat{\mu}}^e U_{\mu,-\hat{\mu}}^{ef}
 \biggr]
   \chi_0^f \chibar_x^b \chi_x^b \rangle
\nn \\
 & - \left( 2 K r i \displaystyle \frac{1}{2^3 \sqrt{3}} \right)^2
   \langle \chibar_0^a \chi_0^a \chibar_0^c \sum_{\alpha,\beta=\pm} \sum_{\mu \neq \nu} (\eta_{\mu\nu,0})^2 
   \biggl[
    (\mathcal{W}_{\alpha\mu\beta\nu,0}^{(2)})^{cd}
    \chi_{\alpha\hat{\mu}\beta\hat{\nu}}^d \chibar_{\alpha\hat{\mu}\beta\hat{\nu}}^e 
    (\mathcal{W}_{\alpha\mu\beta\nu,0}^{(2) \dagger})^{ef}
   \biggr]
   \chi_0^f \chibar_x^b \chi_x^b \rangle 
\ ,
\label{HPE-HoelTwo}   
\end{align}
where $\mathcal{W}_{\alpha\mu\beta\nu,0}^{(2)}$ is defined in Table.~\ref{Table:U2}. 
Note that we consider only connected diagrams in Fig.~\ref{Two-H}.
By integrating out the link variables in the strong-coupling limit,
it is simplified as
\begin{align}
\mathcal{S}(0,x)  \equiv
 \langle \chibar_0^a\chi_0^a \chibar_x^b \chi_x^b \rangle
 = - \delta_{0x} N_c &+ K^2 \sum_{\pm \mu} \langle 
 \chi_{\hat{\mu}}^a \chibar_{\hat{\mu}}^a 
  \chibar_x^b \chi_x^b \rangle
\nn \\
 & + \left( 2 K r i \displaystyle \frac{1}{2^3 \sqrt{3}} \right)^2
  \sum_{\substack{\pm \mu,\pm \nu \\ (\mu \neq \nu)}} 
  \langle \chi_{\hat{\mu}+\hat{\nu}}^a \chibar_{\hat{\mu}+\hat{\nu}}^a 
  \chibar_x^b \chi_x^b \rangle 
\ .
\label{Eq:HPE-Hoelbling2}
\end{align}
Then the self-consistent equation for $\mathcal{S}$ is given in the 
momentum space as
\begin{align}
\mathcal{S}(p) &=
 - N_c +  \biggl[- K^2 \sum_\mu \left( e^{-ip_\mu} + e^{ip_\mu} \right)
\nn \\  
  &+ \left( 2 K r \displaystyle \frac{1}{2^3 \sqrt{3}} \right)^2
  \sum_{\mu \neq \nu} \left( e^{-i(p_\mu+p_\nu)} + e^{i(p_\mu+p_\nu)}
   + e^{-i(p_\mu-p_\nu)} + e^{i(p_\mu-p_\nu)} \right)\biggr] \mathcal{S}(p)
\ .
\label{HPE-HoelSelf2}
\end{align}
We finally obtain the meson propagator as
\begin{equation}
\mathcal{S}(p) = N_c \biggl[ - 2 K^2 \sum_\mu \cos p_\mu 
+ 4 \left( 2 K r \displaystyle \frac{1}{2^3 \sqrt{3}} \right)^2
 \sum_{\mu \neq \nu} \cos p_\mu \cos p_\nu - 1 \biggr]^{-1}
\ .
\label{MP}
\end{equation}
Here the pole of $\mathcal{S}(p)$ should give meson mass.
Since $\chi$ is an one-component fermion,
it may seem to be difficult to find the pion excitation from Eq.~(\ref{MP}).
However, as we discussed,  $\gamma_{5}$ in the staggered fermion 
is given by $\epsilon_{x}=(-1)^{x_{1}+...+x_{4}}$ and the pion 
operator is given by $\pi_{x}=\bar{\chi}_{x}i\epsilon_{x}\chi_{x}$. 
We therefore identify momentum of pion by measuring it from a shifted origin 
$p=(\pi,\pi,\pi,\pi)$. Here we set $p=(i m_\pi a + \pi, \pi,\pi,\pi)$ for 
$1/\mathcal{S}(p)=0$ in Eq.~(\ref{MP}). Then we derive the pion mass $m_\pi$ as
\begin{align}
\cosh(m_\pi a) &= 1 + \displaystyle \frac{1-16K^2}{6K^2}
\ ,
\label{HPE-Hoelpi}
\end{align}
where we again set $r=2\sqrt{2}$ for simplicity.
In this result, the pion mass becomes zero at $|K|=1/4$, 
and tachyonic in the range $\mid{K}\mid > 1/4$. 
It implies that there occurs a phase transition between
parity-symmetric and parity-broken phases at $|K|=1/4$, 
which is consistent with the result from the one-point function 
in Eq.~(\ref{cond}). 
We note that the massless pion at the phase boundary is 
consistent with the scenario of second-order transition.
We can also derive the sigma meson mass by substituting
$p=(i m_\pi a, 0, 0, 0)$ for $1/\mathcal{S}(p)=0$ in Eq.~(\ref{MP}) as
\begin{align}
\cosh(m_\sigma a) &= 1 + \displaystyle \frac{1}{2K^2}
\ .
\label{HPE-Hoelsigma}
\end{align}

\subsection{Adams type}
\label{subsec:HPE-A}
\begin{figure}
 \begin{center}
  \includegraphics[height=6cm]{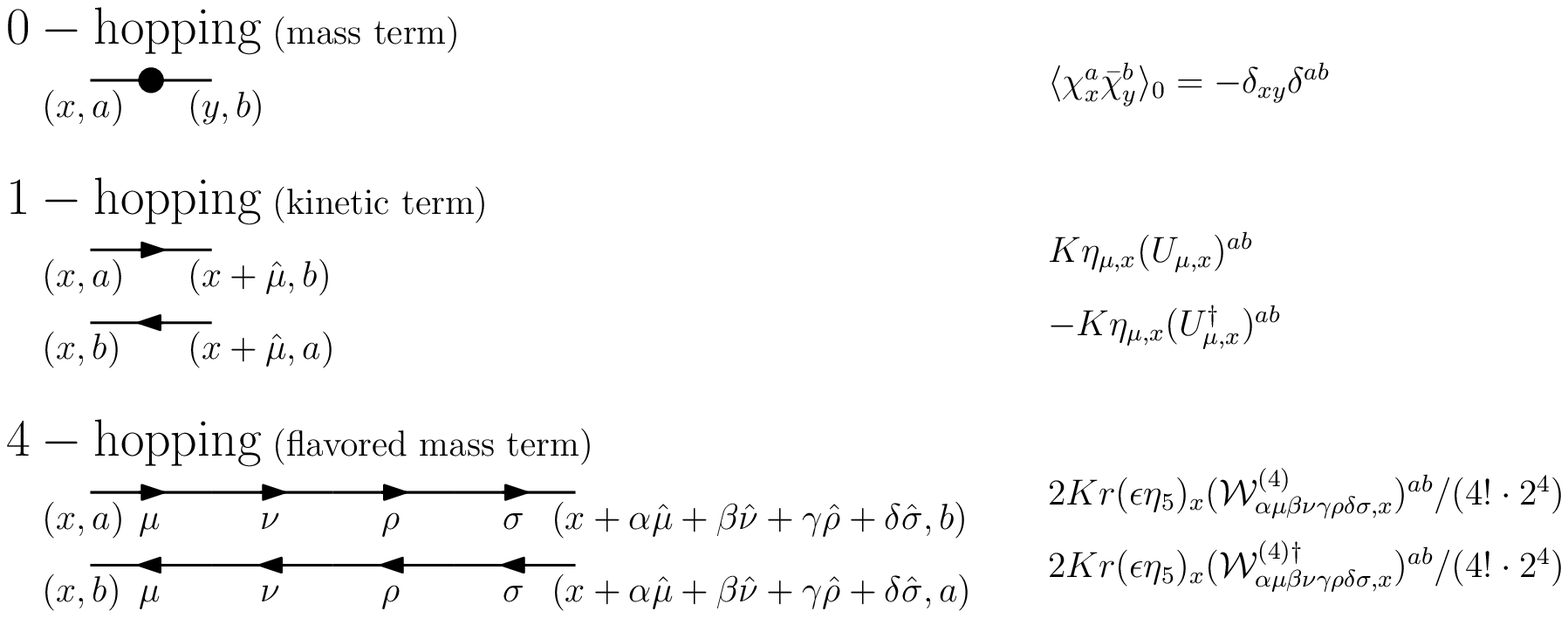}\,\,\,\,
   \end{center}
 \caption{Feynman rules for the HPE with the Adams fermion. $a$ and $b$ stand for the color indices. We show the concrete forms of $\mathcal{W}_{\alpha\mu\beta\nu\gamma\rho\delta\sigma,x}^{(4)}$ in Table \ref{Table:U4} of Appendix \ref{AdamsEff}.}
 \label{FR-A}
\end{figure}

\begin{figure}
 \begin{center}
  \includegraphics[height=3cm]{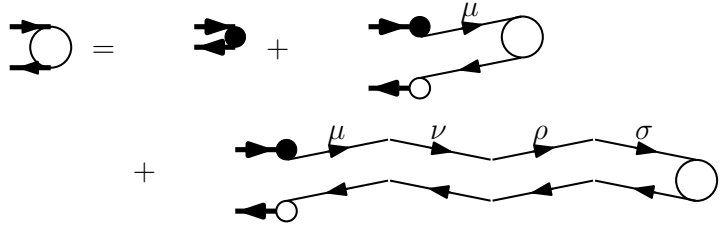}
 \end{center}
 \caption{Feynman diagram for mesonic one-point functions in the $\mathcal{O}(K^{3})$
 self-consistent equations of HPE with the Adams fermion. 
 There is a 4-hopping fundamental diagram,
 which is peculiar to this fermion.}
 \label{One-A}
\end{figure}

We investigate the parity-phase structure for the Adams-type staggered-fermion 
by using $\mathcal{O}(K^{3})$ self-consistent equations in hopping parameter expansion.
The approach is basically parallel to that of Hoelbling type. 
We just need to consider Feynman diagrams for this case.  
The action Eq.~(\ref{AdamsS}) is rewritten by redefining 
$\chi \rightarrow \sqrt{2K} \chi$ with $K=1/[2(M+r)]$ as,
\begin{equation}
S =
\sum_{x} \chibar_x \chi_x + 2K \sum_{x, y} \chibar_{x}(\eta_{\mu}  
D_\mu )_{xy}\chi_{y}+ 2 K r \sum_{x,y} \chibar_{x} 
(M_{A})_{xy} \chi_{y}
\ ,
\label{HPE-Adams}
\end{equation}
where $M_{A}$ is given in Eq.~(\ref{AdamsM}).
In Fig.~\ref{FR-A}, the Feynman rules in the HPE 
for this fermion are depicted.
%%%%%%%%
%%%%%%%%%
First, we derive meson condensates from the one-point function 
$\mathcal{M}_{x}=\bar{\chi}_{x}\chi_{x}$.
The equation for the one-point function is 
obtained as shown in Fig. \ref{One-A},
\begin{align}
- \Sigma_x
 & \equiv
 - \langle \mathcal{M}_x \rangle
\nn \\
 & =
 - \langle \mathcal{M}_x \rangle_0 + 2K^2 \sum_\mu \Sigma_x \Sigma_{x+\hat{\mu}}
 - 2 \cdot \displaystyle \frac {1}{(4!)^2 \cdot 2^3} (Kr)^2 
 \sum_{\mu \neq \nu \neq \rho \neq \sigma} 
 \Sigma_x \Sigma_{x+\hat{\mu}+\hat{\nu}+\hat{\rho}+\hat{\sigma}}
\ .
\label{Eq:HPE-Adams1}
\end{align}
%%%%%%%%
%%%%%%%%%
We substitute $\Sigma_x=\sigma_x +i \epsilon_x \pi_x$ 
for $\Sigma_{x}$ in Eq.~(\ref{Eq:HPE-Adams1}) and obtain the 
self-consistent equation 
\begin{equation}
- \left( \sigma +i \epsilon_x \pi \right)=
 -1 + 2K^2 \cdot 4 \left( \sigma^2 + \pi^2 \right)
 - 2 \cdot \displaystyle \frac {1}{(4!)^2 \cdot 2^3} (Kr)^2 \cdot 4! 
 \left( \sigma +i \epsilon_x \pi \right)^2 
\ .
\label{HPE-AdamsSelf1}
\end{equation}
From this, we obtain $- \sigma = -1 + 16 K^2 \pi^2$ and 
$- i \pi = - 8K^2 \cdot 2 i \sigma \pi$. Here we have set $r=16\sqrt{3}$ 
to make the equation simple. We again have two solutions: 
For $\pi=0$, we have a trivial solution $\sigma=1$. 
For $\pi \neq 0$, we have a non-trivial solution as
\begin{equation}
\sigma = \displaystyle \frac{1}{16K^2},\,\,\,\,\,\,\,\,\,\,\,\,\,\,
\pi = \pm \sqrt{ \displaystyle \frac{1}{16K^2} \left( 1- \displaystyle 
\frac{1}{16K^2} \right) }
\ .
\label{condA}
\end{equation}
It indicates that parity-broken phase 
appears in the range of the hopping parameter as $\mid{K}\mid > 1/4$ 
or equivalently $-16\sqrt{3}-2<M<-16\sqrt{3}+2$.

\begin{figure}
 \begin{center}
  \includegraphics[height=4cm]{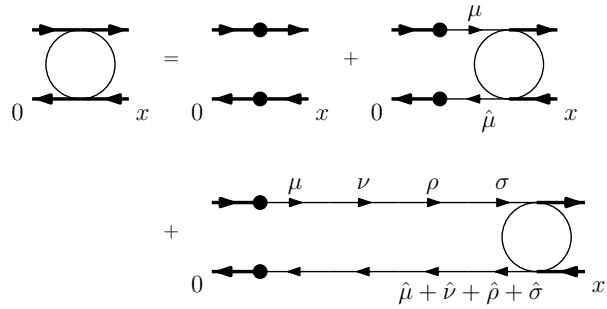}
 \end{center}
 \caption{Feynman diagram for mesonic two-point functions for
$ \mathcal{O}(K^{3})$ self-consistent equations of HPE with the Adams fermion.}
 \label{Two-A}
\end{figure}

Next, we discuss the meson mass from a two-point function of the meson operator 
$\mathcal{S}(0,x)\equiv\langle\mathcal{M}_{0}\mathcal{M}_{x}
\rangle$. 
From Fig. \ref{Two-A} we derive the following equation 
for two-point functions,
%%%%%%%%
%%%%%%%%%
\begin{align}
\mathcal{S}(0,x) 
 = &\langle \chibar_0^a\chi_0^a \chibar_x^b \chi_x^b \rangle
\nn \\
=&- \delta_{0x} N_c
\nn \\
 &-K^2 \langle \chibar_0^a \chi_0^a \chibar_0^c (\eta_{\mu,0})^2 
 \biggl[
   U_{\mu,0}^{cd} \chi_{\hat{\mu}}^d \chibar_{\hat{\mu}}^e 
   (U_{\mu,0}^{\dagger})^{ef}+ (U_{\mu,-\hat{\mu}}^{\dagger})^{cd} 
   \chi_{-\hat{\mu}}^d \chibar_{-\hat{\mu}}^e U_{\mu,-\hat{\mu}}^{ef}\biggr]
   \chi_0^f \chibar_x^b \chi_x^b \rangle
\nn \\
  & + \left( 2 K r \epsilon \eta_5 \displaystyle \frac{1}{4! \cdot 2^4} \right)^2
   \langle \chibar_0^a \chi_0^a \chibar_0^c 
   \sum_{\alpha,\beta,\gamma,\delta=\pm} \sum_{\mu \neq \nu \neq \rho \neq \sigma} 
   \biggl[
   \left( \mathcal{W}_{\alpha\mu\beta\nu\gamma\rho\delta\sigma,0}^{(4)} \right)^{cd}
   \chi_{\alpha\hat{\mu}\beta\hat{\nu}\gamma\hat{\rho}\delta\hat{\sigma}}^d 
    \chibar_{\alpha\hat{\mu}\beta\hat{\nu}\gamma\hat{\rho}\delta\hat{\sigma}}^e 
\nn \\
 & \times
\left( \mathcal{W}_{\alpha\mu\beta\nu\gamma\rho\delta\sigma,0}^{(4) \dagger} \right)^{ef}
 \biggr]
 \chi_0^f \chibar_x^b \chi_x^b \rangle
\ ,
\label{HPE-AdamsTwo}   
\end{align}
where $\mathcal{W}_{\alpha\mu\beta\nu\gamma\rho\delta\sigma,x}^{(4)}$ is defined 
in Table.~\ref{Table:U4} of Appendix \ref{AdamsEff}.
By integrating out the link variables in the strong-coupling limit,
it is simplified as
\begin{align}
\mathcal{S}(0,x)  \equiv
 \langle \chibar_0^a\chi_0^a \chibar_x^b \chi_x^b \rangle
 = &- \delta_{0x} N_c + K^2 
  \sum_{\pm \mu} \langle \chi_{\hat{\mu}}^a \chibar_{\hat{\mu}}^a 
  \chibar_x^b \chi_x^b \rangle
\nn \\
 & - \left( 2 K r \displaystyle \frac{1}{4! \cdot 2^4} \right)^2
  \sum_{\substack{\pm \mu, \pm \nu, \pm \rho, 
  \pm \sigma \\ (\mu \neq \nu \neq \rho \neq \sigma)}} 
  \langle \chi_{\hat{\mu}+\hat{\nu}+\hat{\rho}
  +\hat{\sigma}}^a \chibar_{\hat{\mu}+\hat{\nu}+\hat{\rho}
  +\hat{\sigma}}^a \chibar_x^b \chi_x^b \rangle 
\ .
\label{Eq:HPE-Adams2}
\end{align}
Then the self-consistent equation for $\mathcal{S}$ is given in the 
momentum space as
\begin{align}
\mathcal{S}(p) &=
- N_c + 
 \biggl[- K^2 \sum_\mu \left( e^{-ip_\mu} + e^{ip_\mu} \right)
\nn \\  
 &+ 
  \left( 2 K r \displaystyle \frac{1}{4! \cdot 2^4} \right)^2
    \sum_{\alpha,\beta,\gamma,\delta=\pm} \sum_{\mu \neq \nu \neq \rho \neq \sigma} 
   e^{+i(\alpha p_\mu + \beta p_\nu + \gamma p_\rho + \delta p_\sigma)} 
 \biggr] 
\ .
\label{HPE-AdamsSelf2}
\end{align}
We finally obtain the meson propagator as
\begin{equation}
S(p) = N_c
\biggl[ - 2 K^2 \sum_\mu \cos p_\mu 
+ 16 \left( 2 K r \displaystyle \frac{1}{4! \cdot 2^4} \right)^2
 \sum_{\mu \neq \nu \neq \rho \neq \sigma} 
 \cos p_\mu \cos p_\nu \cos p_\rho \cos p_\sigma- 1 \biggr]^{-1}
\ .
\label{MP2}
\end{equation}
Here we set $p=(i m_\pi a + \pi, \pi,\pi,\pi)$ for 
$1/\mathcal{S}(p)=0$ in Eq.~(\ref{MP2}), 
which gives the pion mass $m_\pi$ as
\begin{align}
\cosh(m_\pi a) &= 1 + \displaystyle \frac{1-16K^2}{10K^2}
\ ,
\label{HPE-Adamspi}
\end{align}
where we again set $r=16\sqrt{3}$ for simplicity.
Here the pion mass becomes zero at $\mid{K}\mid = 1/4$ and
becomes tachyonic in the range $\mid{K}\mid > 1/4$. It suggests that 
there occurs a second-order phase transition 
between parity-symmetric and broken phases at $|K|=1/4$,  
which is consistent with Eq.~(\ref{condA}).
We can also derive the sigma meson mass by substituting
$p=(i m_\pi a, 0, 0, 0)$ for $1/\mathcal{S}(p)=0$ in Eq.~(\ref{MP2}) as
\begin{align}
\cosh(m_\sigma a) &= 1 + \displaystyle \frac{1}{6K^2}
\ .
\label{HPE-Asigma}
\end{align}

%%%%%%%%   Effective Potential Analysis  %%%%%%%%%%%%%

\section{Effective Potential Analysis}
\label{sec:EPA}

In the previous section, we have investigated the parity-phase structure in hopping parameter expansion.
We found a strong sign of parity-broken phase for $\mid{K}\mid >1/4$.
In order to judge whether the parity-broken phase is realized as a vacuum,
the analysis of the gap solution in the hopping parameter expansion is not enough,
and we need to investigate the effective potential for meson fields.

In this section, we consider the effective potential of meson
fields for $SU(N)$ lattice gauge theory with staggered-Wilson fermions.
In the strong-coupling limit and the large $N$ limit, 
effective action can be exactly derived by integrating the link variables \cite{KaS, AokiP}.
Then, by solving a saddle-point equation, we can investigate a vacuum
and find meson condensations. 
In this section we again begin with the Hoelbling case as exercise, 
and go on to Adams fermion with better discrete symmetry.

\subsection{Hoelbling type}
In the strong-coupling limit we can drop plaquette action.  
Then the partition function for meson fields
$\mathcal{M}_x=(\chibar_x \chi_x)/N$ with the source $J_{x}$ is given by
\begin{align}
Z(J) &= \displaystyle \int \Fint \left[ \chi, \chibar, U \right]
 \exp \left[ N \sum_x J_x \mathcal{M}_x + S_F \right]
\ .
\label{ZJ}
\end{align}
where $S_{F}$ stands for the fermion action. 
Here we have defined $\mathcal{M}_x$ with a $1/N$ factor to ensure it to have
a certain large $N$ limit.
In this case, $S_F$ is the Hoelbling-type 
staggered-Wilson action Eq.~(\ref{HoelS}). $N$ stands for the number of color. 
In the large $N$ limit, we can perform the link integral. 
We here consider the effective action up to $\mathcal{O}(\mathcal{M}^{3})$
for meson field $\mathcal{M}$.
This order corresponds to the $\mathcal{O}(K^{3})$ self-consistent equation 
in the hopping parameter expansion. 

We develop a method to perform the link-variable integral with multi-hopping fermion action terms.
In our method, we perform the link integral by introducing two kinds of link-variable
measures.  Now we formally rewrite the partition function as,
\begin{align}
Z(J) &= \displaystyle \int \Fint \left[ \chi, \chibar \right]
 \exp \left[ \sum_x N \left( J_x + \Mhat \right) \mathcal{M}_x \right]
 \exp \left[ \sum_x N W (\Lambda) \right]
\ ,
\end{align}
where we define $\Mhat=M+2r$ and the last term is expressed as,
\begin{align}
\exp \left[ \sum_x N W (\Lambda) \right]
&=  
\prod_x Z_x
\ , \nn \\
Z_x
&= \displaystyle \int \left( \prod_{\mu \neq \nu} \Fint \left[ U_{\mu,x}, 
U_{\mu,x+\hat{\nu}} \right] \right) 
\exp \left[ - \left( \Tr (VE^\dagger) - \Tr (V^\dagger E) \right) \right]
\ .
\label{WZ}
\end{align}
\label{sec:EPA-H}
$\Lambda$, $E$ and $E^\dagger$ ($V$ and $V^\dagger$) are composites of the fermion field $\chi$ (link variables $U$), which we will explicitly show
later. $W(\Lambda)$ is a function of $\Lambda$, which will be an essential part of
effective potential of meson fields. 

\begin{figure}
 \begin{center}
  \includegraphics[height=5cm]{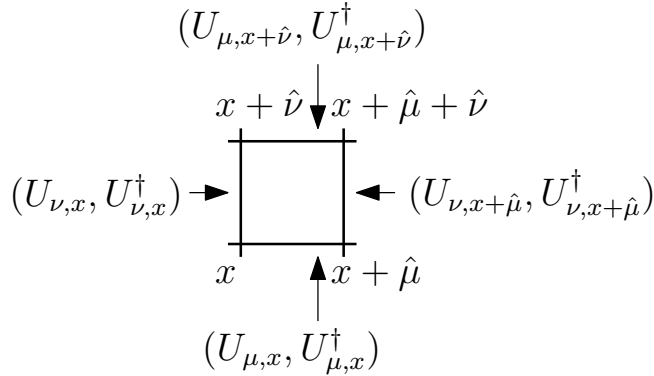}\,\,\,\,
   \end{center}
 \caption{Link variables corresponding to the two kinds of measures in 
 the partition function Eq. (\ref{WZ}) in a 2 dimensional case.}
 \label{H-2link}
\end{figure}

Now we explain how the integral in Eq.~(\ref{WZ}) can be performed by using two types of 
the link measure. Let us consider two-dimensional cases in Fig.~\ref{H-2link} for simplicity. 
In this case, $U_{\mu,x}$ and $U_{\mu, x+\hat{\nu}} \ (\mu \neq \nu)$ form link variables 
in a square block. We can classify diagrams to $\mathcal{O}(\mathcal{M}^{3})$
into three types: (1) 1-link ($\mu$) + 1-link ($-\mu$) hoppings, (2) 2-link ($\mu,\nu$) 
+ 2-link ($-\nu,-\mu$) hoppings, (3) 2-link ($\mu,\nu$) + 1-link ($-\nu$) + 1-link ($-\mu$) hoppings. 
The 1-link hopping comes from the usual staggered kinetic term while
the 2-link hopping from the flavored-mass term.
(1) and (2) are $\mathcal{O}(\mathcal{M}^{2})$ while (3) is $\mathcal{O}(\mathcal{M}^{3})$.
Since one square block contains all the three diagrams,
we can derive the effective potential by integrating link variables per each block. We note that 
$\mathcal{O}(\mathcal{M}^{3})$ diagrams cancel between one another, which is
consistent with the HPE calculations. 
We can also avoid double-counting by adjusting
factors for one-link and two-link terms as we will show in Eq.~(\ref{Lambda}).
%%%%%%%%
%%%%%%%%%

\begin{figure}
 \begin{center}
  \includegraphics[height=7cm]{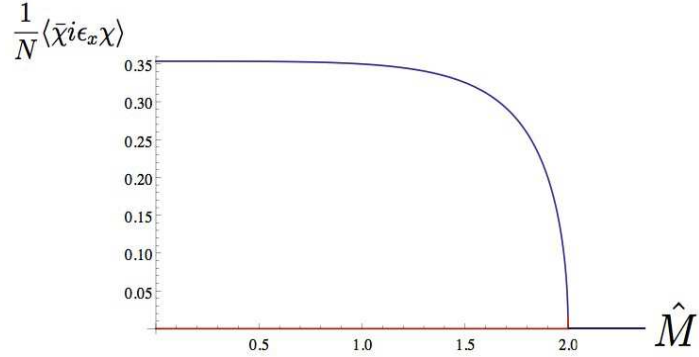}\,\,\,\,
   \end{center}
 \caption{The pion condensate undergoes second-order phase transition.}
 \label{trans1}
\end{figure}

In this method, we need to define sets of link variables and fermion bilinears 
as $V$ and $E$ in Eq.~(\ref{WZ}) : $V$ and $E$  
are matrices including components corresponding to $1$- and 
$2$-link terms. We call a space spanned by these matrices ``hopping space". 
Here we define $a,b$ and $\alpha, \beta$ as color and hopping 
space indices respectively. We also denote $\Tr$ as trace for color 
and the hopping space. Explicit forms of $V$ and $E$ are given by
\begin{align}
V^{ab}_{\alpha \beta}
&= \mathrm{diag} \left(
V_1^{ab}, V_2^{ab}, V_3^{ab}
\right)
\ ,
\end{align}
with
\begin{align}
V_1
&=\mathrm{diag} \left(U_{\mu,x} \right)
\ \nn \\
&\equiv \mathrm{diag} \underbrace{ \left( U_{1,x}, U_{2,x}, \cdots, U_{4,x} \right) }_4
\ , \\
V_2
&=\mathrm{diag} \left(
U_{\mu,x} U_{\nu,x+\hat{\mu}} 
\right)
\ \nn \\
&\equiv \mathrm{diag} \underbrace{ \left(U_{1,x} U_{2,x+\hat{1}}, U_{1,x} U_{3,x+\hat{1}} , \cdots, U_{4,x} U_{3,x+\hat{4}}\right) }_{12}
\ , \\
V_3
&=\mathrm{diag} \left(
U_{\mu,x+\hat{\nu}} U_{\nu,x+\hat{\mu}}^\dagger 
\right)
\ \nn \\
&\equiv \mathrm{diag} \underbrace{ \left(U_{1,x+\hat{2}} U_{2,x+\hat{1}}^\dagger, U_{1,x+\hat{3}} U_{3,x+\hat{1}}^\dagger , \cdots, U_{4,x+\hat{3}} U_{3,x+\hat{4}}^\dagger \right) }_{12}
\ ,
\end{align}
\begin{align}
E^{ab}_{\alpha \beta}
&= \mathrm{diag} \left(
E_1^{ab}, E_2^{ab}, E_3^{ab}
\right)
\ ,
\end{align}
and
\begin{align}
E_1
&= \mathrm{diag} \left(
D_{1,\mu}
\right)
\ \nn \\
&\equiv \mathrm{diag} \underbrace{ \left(
D_{1,1}, D_{1,2}, \cdots, D_{1,4}
\right) }_4
\ , \\
E_i
&= \mathrm{diag} \left(
D_{i,\mu\nu}
\right)
\ \nn \\
&\equiv 
\mathrm{diag} \underbrace{ \left(
D_{i,12}, D_{i,13}, \cdots, D_{i,43}
\right) }_{12}
, \ (i=2,3)
\ ,
\end{align}
where we define the operator $D$ as the fermion bilinears,
\begin{align}
\left( D_{1,\mu}^\dagger \right)^{ab} &= 
\displaystyle \frac {1}{2} \eta_{\mu,x} \chibar_x^a \chi_{x+\hat{\mu}}^b 
\ , \quad
\left( D_{1,\mu} \right)^{ab} =
\displaystyle \frac {1}{2} \eta_{\mu,x} \chibar_{x+\hat{\mu}}^a \chi_x^b
\ , \\
\left( D_{2,\mu \nu}^\dagger \right)^{ab} &= 
\displaystyle \frac {ir}{2^3 \sqrt{3}} \eta_{\mu \nu,x}
\chibar_x^a \chi_{x+\hat{\mu}+\hat{\nu}}^b 
\ , \quad
\left( D_{2,\mu \nu} \right)^{ab} = 
\displaystyle \frac {ir}{2^3 \sqrt{3}} \eta_{\mu \nu,x} 
\chibar_{x+\hat{\mu}+\hat{\nu}}^a \chi_x^b 
\ , \\
\left( D_{3,\mu \nu}^\dagger \right)^{ab} &= 
\displaystyle \frac {ir}{2^3 \sqrt{3}} \eta_{\mu \nu,x+\hat{\nu}} 
\chibar_{x+\hat{\nu}}^a \chi_{x+\hat{\mu}}^b 
\ , \quad
\left( D_{3,\mu \nu} \right)^{ab} = 
\displaystyle \frac {ir}{2^3 \sqrt{3}} \eta_{\mu \nu,x+\hat{\nu}} 
\chibar_{x+\hat{\mu}}^a \chi_{x+\hat{\nu}}^b 
\ .
\end{align}
Here $V_{1}$ and $E_{1}$ are $4\times 4$ diagonal matrices 
while $V_{i}$ and $E_{i}$ $(i=2,3)$ are $12\times 12$ diagonal matrices.
Now we have prepared to obtain $W(\Lambda)$.
By using the relation $U^{\dagger}U=1$, we obtain 
the Schwinger-Dyson equation,
\begin{align}
\displaystyle \frac{\partial^2 Z_x}{\partial 
E^{ab}_{\alpha \beta} \partial 
\left( E^\dagger \right)^{bc}_{\beta \gamma}}
&= - \delta_{ca} \delta^{\alpha \gamma} Z_x
\ .
\end{align}
$W$ should be a function of a gauge-invariant quantities as follows.
\begin{align}
\Lambda^{ab}_{\alpha \beta} &= \displaystyle
 \frac {1}{N^2} \left( E^\dagger E \right)^{ab}_{\alpha \beta} 
\ .
\end{align}
We can solve the Schwinger-Dyson equation analytically 
and derive $W$ as a function of $\Lambda$,
\begin{align}
W(\Lambda)
&= \Tr \left[ \left( 1-4 \Lambda \right)^{1/2} - 1 
- \ln  \left[ \displaystyle \frac {1+\left( 1-4 \Lambda \right)^{1/2}}{2} 
\right] \right]
\ .
\end{align}
We here perform trace for colors and hopping spaces.
\begin{align}
\sum_x W(\Lambda)
&= - \sum_x \left[ \left( 1-4 \Lambda_x \right)^{1/2} 
- 1 - \ln  \left[ \displaystyle \frac {1+
\left( 1-4 \Lambda_x \right)^{1/2}}{2} \right] \right]
\ .
\end{align}
Finally we obtain a concrete form of $\Lambda$ as
\begin{align}
\Lambda_x
&=
\displaystyle \frac {1}{8}  
\left[ 
 \sum_\mu \mathcal{M}_x \mathcal{M}_{x+\hat{\mu}}
 + \displaystyle \frac {1}{3}  
   \sum_{\mu \neq \nu} \mathcal{M}_{x+\hat{\mu}} \mathcal{M}_{x+\hat{\mu}+\hat{\nu}}
\right]
-\left( \displaystyle \frac {r}{2^3 \sqrt{3}} \right)^2 
\sum_{\mu \neq \nu} \left( \mathcal{M}_x \mathcal{M}_{x+\hat{\mu}+\hat{\nu}} 
+ \mathcal{M}_{x+\hat{\nu}} \mathcal{M}_{x+\hat{\mu}}\right)
\ .
\label{Lambda}
\end{align}
The first and second terms correspond to contribution from $D_{1,\mu},D_{1,\mu}^\dagger$,
and the third and forth terms correspond to contribution 
from $D_{i,\mu},D_{i,\mu}^\dagger \ (i=2,3)$.
Now we again set $r=2\sqrt{2}$ to compare the result to
that of the hopping parameter expansion in Sec.~\ref{sec:HPE}.
We need to change the fermion measure to the meson-field 
measure as
\begin{align}
\displaystyle \int \Fint \left[ \chi, \chibar \right] 
&= \displaystyle \int \Fint \mathcal{M} \exp \left[ -N \sum_x \ln \mathcal{M}_x \right] 
\ .
\end{align}
Then the effective partition function for the meson field is given by
\begin{align}
Z(J)
&= \displaystyle \int \Fint \mathcal{M} \exp \left[ N \left( \sum_x J_x \mathcal{M}_x 
+ S_{\mathrm{eff}}(\mathcal{M}) \right) \right] 
\ , \\
S_{\mathrm{eff}}(\mathcal{M})
&= \sum_x \left( \Mhat \mathcal{M}_x - \ln \mathcal{M}_x \right) + \sum_x W(\Lambda)
\ ,
\label{EAc1}
\end{align}
where we denote $\Mhat$ as the shifted mass parameter $\Mhat = M+2r$.
The partition function with $J=0$ in the large $N$ limit is 
reduced to the integrant for the saddle-point values of the meson fields,
\begin{align}
Z(J=0)
&= \displaystyle \int \Fint \mathcal{M} \exp \left[ N S_{\mathrm{eff}}(\mathcal{M}) \right]
\ \nn \\
& \sim \exp \left[ N S_{\mathrm{eff}}(\bar{\mathcal{M}}) \right], 
\quad \left( N \rightarrow \infty \right)
\ .
\end{align}
Now we consider pion condensate.
For now, we consider only scalar $\sigma$ and pseudo-scalar 
$\pi$ fields as
\begin{align}
\bar{\mathcal{M}}_{x}
&= \sigma + i \epsilon_x \pi 
\ ,
\\
&= \Sigma e^{i\epsilon_x \theta}
\ .
\end{align}
By substituting this form of the meson field into the Eq.~(\ref{EAc1}),
we derive the effective action for the $\Sigma$ and $\theta$,
\begin{align}
S_{\mathrm{eff}}(\bar{\mathcal{M}})
&= \Mhat \sum_x \Sigma \cos \theta - \sum_x \ln \Sigma
\ \nn \\
&- \sum_x \left[ \left( 1- 4 \cdot 2 \Sigma^2 \sin^2 \theta \right)^{1/2} 
 - \ln \left[ \displaystyle \frac{1+\left( 1- 4 \cdot 2 \Sigma^2 \sin^2 
 \theta \right)^{1/2}}{2} \right] \right]
\ .
\end{align}
We ignore the irrelevant constant.
From the translational invariance, we factorize the 4-dimensional 
volume $V_4$ from the effective action as 
$S_{\mathrm{eff}}(\bar{M})=-V_4 V_\mathrm{eff}(\Sigma,\theta)$.
Then the effective potential $V_{\mathrm{eff}}$ is given by
\begin{align}
V_{\mathrm{eff}}(\Sigma,\theta)
&= - \Mhat \Sigma \cos \theta + \ln \Sigma
\ \nn \\
&+ \left[ \left( 1- 4 \cdot 2 \Sigma^2 \sin^2 \theta \right)^{1/2} 
 - \ln \left[ \displaystyle \frac{1+\left( 1- 4 \cdot 2 \Sigma^2 
 \sin^2 \theta \right)^{1/2}}{2} \right] \right]
\ .
\label{effS1}
\end{align}
Now let us look into the vacuum structure of this effective potential
by solving the saddle-point condition, which are given by
\begin{align}
\displaystyle \frac {\partial V_{\mathrm{eff}}(\Sigma,\theta)}{\partial \Sigma}
&= - \Mhat \cos \theta + \displaystyle \frac {1}{\Sigma}
 - \displaystyle \frac {8 \Sigma \sin^2 \theta}{1+ 
 \left( 1- 4 \cdot 2 \Sigma^2 \sin^2 \theta \right)^{1/2}}
\quad 
=0
\ , \\
\displaystyle \frac {\partial V_{\mathrm{eff}}(\Sigma,\theta)}{\partial \theta}
&= \Sigma \sin \theta \left[ 
 \Mhat - \displaystyle \frac {8 \Sigma \cos \theta}{1
 + \left( 1- 4 \cdot 2 \Sigma^2 \sin^2 \theta \right)^{1/2}}
 \right]
\quad 
=0
\ .
\end{align}
Here we find two types of solutions for these equations
depending on whether $\theta$ is zero or nonzero:
For a trivial solution $\theta=0$, 
we have $\Sigma=1/\Mhat$.
For $\theta \neq 0$, the stationary conditions are written as
\begin{align}
&\Mhat \Sigma - \cos \theta =0
\ , \\
&1 - \displaystyle \frac {8 \Sigma^2}{1+ \left( 
1- 4 \cdot 2 \Sigma^2 \sin^2 \theta \right)^{1/2}}
=0
\ .
\end{align}
Then, we find a solution for $\theta\not= 0$ as
\begin{align}
\Sigma &= \bar{\Sigma}= \sqrt{ \displaystyle \frac {1}{8-\Mhat^2}}
\ , \\
\sin^2 \theta &= \sin^2 \bar{\theta}= \displaystyle \frac {2 ( 4-\Mhat^2 ) }{8-\Mhat^2}
\ .
\end{align}
Now we need to figure out which solution is realized as the vacuum of 
the theory by comparing the potentials for the two solutions.
We easily show for $\Mhat^2 < 4$,
\begin{equation}
V_{\mathrm{eff}}(1/\Mhat,0)-V_{\mathrm{eff}}(\bar{\Sigma},\bar{\theta})>0
\ .
\end{equation}
while $V_{\mathrm{eff}}(1/\Mhat,0)-V_{\mathrm{eff}}(\bar{\Sigma},\bar{\theta})<0$
for $\Mhat^2 > 4$.
Thus the vacuum of the strong-coupling QCD with 
the Hoelbling-type staggered-Wilson fermion is given by the following:
For $\Mhat^2 > 4$ or equivalently $M > -4\sqrt{2}+2,\,\, M<-4\sqrt{2}-2$,
there is only the chiral condensate as
\begin{align}
\displaystyle \frac {1}{N} \langle \chibar \chi \rangle 
&= \Sigma \cos \theta \quad = \displaystyle \frac {1}{\Mhat}
\ , \\
\displaystyle \frac {1}{N} \langle \chibar i \epsilon_x \chi \rangle 
&= \Sigma \sin\theta \quad = 0
\ .
\end{align}
For $\Mhat^2 < 4$ or equivalently $-4\sqrt{2}-2<M<-4\sqrt{2}+2$,
there is pion condensate which breaks parity symmetry spontaneously.
\begin{align}
\displaystyle \frac {1}{N} \langle \chibar \chi \rangle &=
\bar{\Sigma} \cos \bar{\theta} \quad = \displaystyle \frac {\Mhat}{8-\Mhat^2}
\ , \\
\displaystyle \frac {1}{N} \langle \chibar i \epsilon_x \chi \rangle 
&= \bar{\Sigma} \sin \bar{\theta} \quad =  
\pm \displaystyle \frac {\sqrt{ 2(4-\Mhat^2)}}{8-\Mhat^2}
\ .
\label{picond}
\end{align}
The sign of the pion 
 condensate Eq.~(\ref{picond}) reflects the $Z_2$ parity symmetry of the theory.
The critical mass parameter $M_{c}=-4\sqrt{2}\pm 2$ and the range for the Aoki phase
$-4\sqrt{2}-2<M<-4\sqrt{2}+2$ is consistent with those of the hopping 
parameter expansion shown below Eq.~(\ref{cond}). 
These results strongly suggest the existence of the parity-broken phase in the lattice
QCD although it is just a strong-coupling limit.
Figure~\ref{trans1} shows  that the phase transition is second-order.
%%%%%%%%
%%%%%%%%%

We can also derive mass spectrum of mesons by expanding the 
effective action Eq.~(\ref{EAc1}) up to the quadratic terms of  
the meson-excitation field $\Pi_x=\mathcal{M}_x-\bar{\mathcal{M}}_x$.
Since we are interested in the chiral limit which is taken from 
the parity-symmetric phase to the critical line, we here concentrate on 
the pion mass in the parity-symmetric phase. For the parity-symmetric 
phase ($\Mhat^{2}>4$), the quadratic part of the effective action is given by
\begin{align}
S_\mathrm{eff}(\mathcal{M}) - S_\mathrm{eff}(\bar{\mathcal{M}})
&= \sum_{x,y} S_\mathrm{eff}^{(2)} (x,y) \Pi_x \Pi_y
\nonumber\\
&= \int_{-\pi}^\pi \displaystyle \frac{d^4p}{(2\pi)^4} 
 \Pi(-p) \mathcal{D} \Pi(p)
\ ,
\end{align}
where $\Pi(p)$ is the Fourier component of $\Pi_x$,
and
\begin{align}
\mathcal{D}
&= \displaystyle \frac {1}{2\Sigma^2} 
 + \left[ \displaystyle \frac{1}{4} \sum_\mu \cos p_\mu 
  - \displaystyle \frac{1}{24} 
    \sum_{\mu \neq \nu} \left( \cos p_{\mu + \nu} + \cos p_{\mu - \nu} \right) 
\right] 
\ ,
\label{HpoD}
\end{align}
with $p_{\mu\pm\nu}\equiv p_{\mu}\pm p_{\nu}$.
Then we obtain the pion mass by solving $\mathcal{D}=0$ at $p=(i m_\pi a + \pi, \pi,\pi,\pi)$ as
\begin{equation}
\cosh(m_{\pi}a) = 1+{2\Mhat^{2}-8\over{3}}
\ .
\label{HpiM}
\end{equation}
By using the definition $K=1/2\Mhat$ with $\Mhat=M+2r$ and $r=2\sqrt{2}$,
we find $\cosh(m_{\pi}a)=1+(1-16K^{2})/6K^{2}$, which is
consistent with the result of the hopping parameter 
expansion Eq.~(\ref{HPE-Hoelpi}): The pion mass becomes zero at
the critical mass $\Mhat^{2}=4$, which indicates there occurs a second-order 
phase transition between parity-symmetric and broken phases in the 
strong-coupling limit. By defining quark mass as $m_{q}a=\Mhat-\Mhat_{c}$,
we find PCAC relation near the critical mass as 
\begin{equation}
(m_{\pi}a)^{2} = {16\over{3}}m_{q}a+\mathcal{O}(a^{2})
\ .
\label{HpiM2}
\end{equation}
We can also study a case for non-zero spacial
momenta by considering $p=(iEa+\pi,p_{1}a+\pi,p_{2}a+\pi,p_{3}a+\pi)$ 
in Eq.~(\ref{HpoD}). By using the pion mass Eq.~(\ref{HpiM2}) and re-normalizing 
the Dirac operator as $-{8\over{3}}\mathcal{D}\to\mathcal{D}$, we show that
Eq.~(\ref{HpoD}) results in the Lorentz-covariant form up to $\mathcal{O}(a)$ discretization errors,
\begin{equation}
\mathcal{D}=(E^{2}-{\bf p}^{2}-m_{\pi}^{2})a^{2}+\mathcal{O}(a^{3})
\ ,
\label{HLorentz}
\end{equation}
with ${\bf p}^{2}=p_{1}^{2}+p_{2}^{2}+p_{3}^{2}$.
As we discussed in Sec.~\ref{sec:Sym}, we expected that we may 
find a sign of Lorentz symmetry breaking of the Hoelbling fermion in this study. 
However, we eventually cannot find a disease due to the symmetry breaking 
in the strong-coupling study. We consider that it is because Lorentz symmetry 
breaking appears mainly in the gluon sector as shown in \cite{Steve} and it is difficult to 
find it in the meson sector in this limit. In future work, we may be able to find 
it by including higher order corrections of $1/g^{2}$ and $1/N$. 

We here discuss possibility of other condensation.
For this purpose, we consider a general form 
of the meson field as
\begin{equation} 
\bar{\mathcal{M}}_{x}=\sigma+i\epsilon_{x}\pi+\sum_{\mu}(-1)^{x_{\mu}}v_{\mu}
+\sum_{\mu}i\epsilon_{x}(-1)^{x_{\mu}}a_{\mu}
+\sum_{\mu>\nu}(-1)^{x_{\mu}+x_{\nu}}t_{\mu\nu}
\ ,
\label{GeM}
\end{equation}
where we define the vector, axial-vector and tensor meson fields
as $v_{\mu}$, $a_{\mu}$ and $t_{\mu}$.
We can easily show there is no other condensate by substituting this general 
form Eq.~(\ref{GeM}) into the meson action Eq.~(\ref{EAc1}). 
Thus we conclude that the vacuum we obtained is a true one.

The results in this section suggest that the chiral limit can be taken in Hoelbling-fermion
lattice QCD in a parallel manner to Wilson fermion. However we probably need 
to tune other parameters to restore Lorentz symmetry in lattice QCD with this type.
Therefore, what we can state here is just that, if we succeed to restore Lorentz symmetry by 
parameter tuning, this fermion could be applied to lattice QCD as Wilson fermion.

\subsection{Adams type}
\label{sec:EPA-A}
We next investigate the case for the Adams fermion.
We again consider the effective potential up to $\mathcal{O}(\mathcal{M}^{3})$. 
The derivation is almost the same as the Hoelbling case in Subsec.~\ref{sec:EPA-H}.
The main difference between them is the number of the multi-links. 
The fermion of the Adams type includes the four-hopping terms while the Hoelbling one
has the two-hopping terms.
In the appendix \ref{AdamsEff}, we derive the
effective potential for the Adams-type fermion.
Here we only summarize the results.

In this case, we again set $r=16\sqrt{3}$ to match the result to
that of the hopping parameter expansion in Sec.~\ref{sec:HPE}.
We can derive the effective potential for scalar and pseudo-scalar fields 
by assuming condensation as
$\mathcal{M}_{x} = \Sigma e^{i\epsilon_{x}\theta}$.
We note that the functional form of the effective potential is the same as Eq.~(\ref{effS1}).
By solving saddle-point equations, 
we find that the critical mass is given by $\Mhat^{2}_{c}=4$ or
equivalently $M_c=-16\sqrt{3}\pm2$ with $\Mhat=M+r$ and $r=16\sqrt{3}$.
The vacuum in this case is given by following:
For $\Mhat^2 > 4$ or $M > -16\sqrt{3}+2,\,\, M<-16\sqrt{3}-2$,
there is only the chiral condensate as
\begin{align}
\displaystyle \frac {1}{N} \langle \chibar \chi \rangle 
&= \Sigma \cos \theta \quad = \displaystyle \frac {1}{\Mhat}
\ , \\
\displaystyle \frac {1}{N} \langle \chibar i \epsilon_x \chi \rangle 
&= \Sigma \sin\theta \quad = 0
\ .
\end{align}
For $\Mhat^2 < 4$ or $-16\sqrt{3}-2<M<-16\sqrt{3}+2$,
there emerge the pion condensate which breaks the parity symmetry spontaneously.
\begin{align}
\displaystyle \frac {1}{N} \langle \chibar \chi \rangle &=
\bar{\Sigma} \cos \bar{\theta} \quad = \displaystyle \frac {\Mhat}{8-\Mhat^2}
\ , \\
\displaystyle \frac {1}{N} \langle \chibar i \epsilon_x \chi \rangle 
&= \bar{\Sigma} \sin \bar{\theta} \quad =  
\pm \displaystyle \frac {\sqrt{ 2(4-\Mhat^2)}}{8-\Mhat^2}
\ .
\end{align}
We note the critical mass and the parameter range of the Aoki phase 
are consistent with those of the hopping parameter expansion shown below 
Eq.~(\ref{condA}). This result supports the existence of the parity-broken phase 
in the lattice QCD with the Adams fermion again.
The behavior of pion condensate in this case is also given by
Fig.~\ref{trans1}, which shows the order of phase transition is
second-order. This is consistent with the second-order scenario 
that we can take a chiral limit by a mass parameter tuning.

We derive the pion mass from the quadratic parts of the 
effective potential in a parallel way to the Hoelbling type. 
The pion mass for this case is given by
\begin{equation}
\cosh(m_{\pi}a) = 1+{2\Mhat^{2}-8\over{5}}
\ ,
\label{ApiM}
\end{equation}
for the parity-symmetric phase ($\Mhat^{2}>4$).
By using the definition $K=1/2\Mhat$ with $\Mhat=M+r$ and $r=16\sqrt{3}$,
we find $\cosh(m_{\pi}a)=1+(1-16K^{2})/10K^{2}$ which is
consistent with the result of the hopping parameter 
expansion Eq.~(\ref{HPE-Adamspi}): The pion mass becomes zero at
the critical mass $\Mhat^{2}=4$, which indicates there occurs a second-order 
phase transition between parity-symmetric and broken phases in the 
strong-coupling limit.
The PCAC relation holds near the critical mass also in this case.
\begin{equation}
(m_{\pi}a)^{2} = {16\over{5}}m_{q}a+\mathcal{O}(a^{2})
\ .
\label{ApiM2}
\end{equation}
We can also show that the Lorentz-covariant 
dispersion relation recovers in the continuum limit in the Adams-type formalism as
$\mathcal{D}=(E^{2}-{\bf p}^{2}-m_{\pi}^{2})a^{2}+\mathcal{O}(a^{3})$. It is reasonable that 
Lorentz-symmetric dispersion recovers since Adams fermion has
sufficiently large discrete symmetry for Lorentz symmetry restoration.
We can also argue the possibility of other condensations
in the same way: We can show there is no other condensates 
by substituting a general form of the meson fields (\ref{GeM}) 
into the mesonic action for this case.

All these results indicate that, in Adams-type staggered-Wilson, 
we can take a chiral limit by tuning a mass parameter toward the 
second-order critical line from the parity-symmetric phase.
Since Adams fermion has sufficient discrete symmetry, 
it can be straightforwardly applied to lattice QCD in a
parallel manner to Wilson fermions.

%%%%%%%% Two-flavor case %%%%%%%%%%%%%%%%

\section{Discussion on two-flavor case}
\label{sec:Tf}
%%%%%%%%
\begin{figure}
 \begin{center}
  \includegraphics[height=5cm]{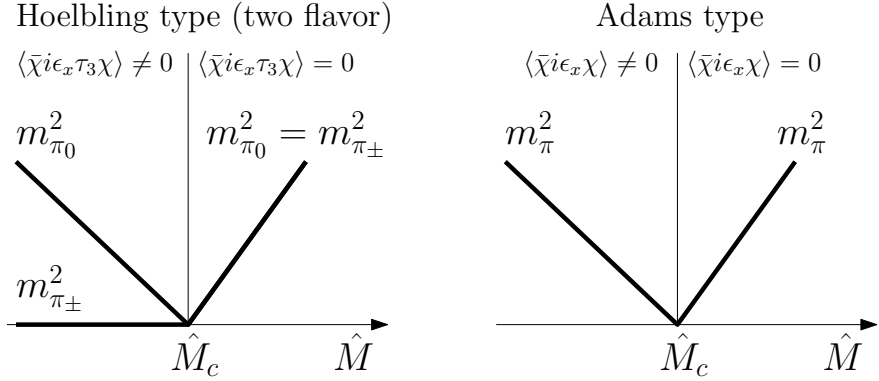}\,\,\,\,
   \end{center}
 \caption{Conjectures of pion mass behaviors as a function of a mass parameter 
 in two-flavor Hoelbling fermions and a single Adams fermion. In both cases PCAC relation
 holds in the parity symmetric phase.}
 \label{Fig:mpi-HA}
\end{figure}
%%%%%%%%%

In this section, we discuss parity-flavor breaking for two-flavor staggered-Wilson
fermions. We first consider two-flavor Hoelbling-type fermion action. 
In this case, except that the low discrete symmetry would require further parameter tuning, 
the situation is quite similar to that of the Wilson fermion \cite{AokiP}.
We here assume that mass parameters for two flavors $M_{f}$ $(f=1,2)$
are equal, which means there is exact $SU(2)$ flavor symmetry. 
The chiral and pion condensates are given by
\begin{align}
{1\over{N}}\langle \bar{\chi}_{f}\chi_{f} \rangle 
&= \Sigma_{f} \cos \theta_{f} = {1\over{\hat{M}_{f}}}
\ ,
\label{2fout1}
\\
{1\over{N}}\langle \bar{\chi}_{f}i\epsilon_{x}\chi_{f} \rangle 
&= \Sigma_{f} \sin \theta_{f}  = 0
\ ,
\label{2fout2}
\end{align}
for $\hat{M}_{f}^{2}\geq 4$ (parity-symmetric phase), while they are given by
\begin{align}
{1\over{N}}\langle \bar{\chi}_{f}\chi_{f} \rangle 
&= \bar{\Sigma}_{f} \cos \bar{\theta}_{f} = {\hat{M}_{f}\over{8-\hat{M}_{f}^{2}}}
\ ,
\\
{1\over{N}}\langle \bar{\chi}_{f}i\epsilon_{x}\chi_{f} \rangle 
&= \bar{\Sigma}_{f} \sin \bar{\theta}_{f} = \pm {\sqrt{2(4-\hat{M}_{f}^{2})}\over{8-\hat{M}_{f}^{2}}}
\ ,
\label{2fcon}
\end{align}
for $\hat{M_{f}}^{2}< 4$ (Aoki phase).
Here we have assumed that only the diagonal condensates in the flavor space
({\em i.e.} neutral condensates) can take finite values.
Here we remind you of $\hat{M}_{f}=M_{f}+2r$ with $r=2\sqrt{2}$.
We first look into the parity-symmetric phase.
Although $SU(2)$ chiral symmetry is explicitly broken due to the 
flavored-mass term, three-massless pions appear on the second-order 
phase boundary due to divergence of correlation length as 
shown in Fig.~\ref{Fig:mpi-HA}. In the parity-broken phase, things 
depend on whether or not $\bar{\theta}_{1}$ and $\bar{\theta}_{2}$ have the same 
sign in Eq.~(\ref{2fcon}). For $\bar{\theta}_{1}=\bar{\theta}_{2}$, 
\begin{align}
&\langle \bar{\chi}i\epsilon_{x}\chi \rangle\not= 0
\ ,
\nonumber\\
&\langle \bar{\chi}i\epsilon_{x}\tau_{i}\chi \rangle=0
\ , \,\,\,\,\,(i=1,2,3)
\ , 
\end{align}
where $\tau_{i}$ is the Pauli matrix and $\chi$ stands a doublet 
$\chi=(\chi_{1},\chi_{2})^T$. For $\bar{\theta}_{1}=-\bar{\theta}_{2}$,
\begin{align}
&\langle \bar{\chi}i\epsilon_{x}\chi \rangle=0
\ ,
\nonumber\\
&\langle \bar{\chi}i\epsilon_{x}\tau_{1}\chi \rangle =0
\ ,
\nonumber\\
&\langle \bar{\chi}i\epsilon_{x}\tau_{2}\chi \rangle =0
\ ,
\nonumber\\
&\langle \bar{\chi}i\epsilon_{x}\tau_{3}\chi \rangle \not=0
\ .
\label{2fVW}
\end{align}
From Vafa-Witten's theorem \cite{VW}, we expect that the latter vacuum 
($\bar{\theta}_{1}=-\bar{\theta}_{2}$) realizes \cite{ACV}. It is also possible to check this
by studying next-leading-order calculation of $1/N$ or $1/g^{2}$ expansions.
If the latter scenario realizes, the flavored pion condensate Eq.~(\ref{2fVW}) breaks 
$SU(2)$ flavor symmetry into its $U(1)$ subgroup as well as parity symmetry
\footnote{In Appendix~\ref{zeroeigen}, we show the relation between an order 
parameter of the phase transition and zero eigenvalues of staggered-Wilson operator, 
as Wilson fermion.}. Thus, in the parity-broken phase, we have two-massless 
pions as NG bosons associated with spontaneous breaking of the flavor symmetry. 
We summarize them in Fig.~\ref{Fig:mpi-HA}. This situation is qualitatively the same 
as the case of Wilson fermion \cite{AokiP} except possibility of further parameter 
tuning to recover Lorentz symmetry.

The Adams-type staggered-Wilson fermion is more fascinating.
It has two flavors for each branch in the first place. 
In this case there is no exact $SU(2)$ flavor symmetry due to taste-mixing in original
staggered fermions. It means that in the parity-broken phase there is no massless 
excitation since there is no continuous symmetry to be broken as Fig.~\ref{Fig:mpi-HA}. 
However the number of massless pions in the chiral limit (on the boundary) depends
on residual discrete flavor symmetry in the pion sector. If the discrete flavor symmetry 
is not sufficient to have a degenerate pion triplet, we have only one massless pion in 
the chiral limit although these three are expected to be degenerate in the continuum limit. 
If the symmetry is large enough, we have three-massless pions on the phase boundary.
The latter is a quite fascinating scenario because we can simulate two-flavor QCD
with a single lattice fermion. It is possible to study it by looking into transfer 
matrix symmetry or chiral Lagrangian potential. Recently Ref.~\cite{Steve} has reported 
that classification of pion operators from the transfer matrix symmetry indicates 
three-degenerate pions. Adams fermion can be a new standard
of lattice fermion in the near future.

%%%%%%%%%%   Summary and Discussion   %%%%%%%%%%

\section{Summary and Discussion}
\label{sec:SD}

In this paper we investigate strong-coupling lattice QCD with staggered-Wilson
fermions, with emphasis on the parity-broken phase (Aoki phase) structure. 
We consider hopping parameter expansion and effective potential 
analysis in the strong-coupling limit. We have shown that the parity-broken 
phase and the second-order phase boundary exist for both Adams-type and 
Hoelbling-type staggered-Wilson fermions, which is consistent with
the second-order scenario for a chiral limit.

In Sec.~\ref{sec:Sym}, we discuss and classify discrete symmetries of
two types of staggered-Wilson fermions. We show that they are invariant
under charge conjugation and parity transformation, the latter of which is 
defined as 4th-shift followed by spatial axis reversal. We also discuss
smaller rotation symmetry in the Hoelbling fermion, which would
require further parameter tuning as shown in \cite{Steve}.

In Sec.~\ref{sec:HPE}, we analyze staggered-Wilson fermions by using 
hopping parameter expansion. From one-point functions of meson fields 
in the expansion, we find that pion condensate becomes nonzero in some 
range of the hopping parameter. From two-point functions, we show that 
square pion mass becomes zero on the boundary and becomes negative in the 
parameter region with nonzero pion condensate. These results suggest 
that there is a parity-broken phase and a second-order phase boundary. 

In Sec.~\ref{sec:EPA}, we study the effective potential for meson fields in 
the strong-coupling limit and large $N$ limit to elaborate the phase structure
in details. Here we develop a method to derive effective 
potential for lattice fermion actions with multiple-hopping terms. 
The gap equations from the effective potential exhibit nonzero pion 
condensate in the same parameter range as the hopping 
parameter expansion. From this analysis, we also 
show that pion becomes massless on the second-order phase boundary, 
and PCAC relation is reproduced around the boundary. If this property 
carries over into the weak-coupling regime, we can take
a chiral limit by tuning a mass parameter in lattice QCD with 
staggered-Wilson fermions as with Wilson fermion. 

In Sec.~\ref{sec:Tf}, we discuss the two-flavor cases.
The situation in two-flavor Hoelbling-type fermions is similar to 
the original Wilson fermion except less rotational symmetry:  
Three-massless pions are expected to
appear on the second-order critical lines, while two of them remain 
massless in the Aoki phase due to the flavored pion condensate. 
However we probably need to care about Lorentz symmetry breaking 
in this case, thus we cannot straightforwardly apply it to two-flavor lattice QCD. 
The Adams-type staggered-Wilson fermion contains 
two flavors in each branch. Although the taste-mixing breaks 
flavor symmetry at finite lattice spacing, it does not necessarily mean
non-degenerate three pions. 
Moreover SU(2) flavor symmetry should recover in the 
continuum limit at least, and three-massless pions emerge 
if we take a chiral and continuum limit. 
In this case, there is no rotational symmetry breaking, and the 
hypercubic symmetry will recover in the continuum limit. 
We can thus perform two-flavor QCD simulations with Adams-type 
staggered-Wilson fermion more efficiently than usual.

All of these results shows new possibilities of lattice fermion formulations.
In particular, the Adams fermion can be straightforwardly applied to 2-flavor lattice 
QCD since it does not require any other fine-tuning and automatically has two flavors.  
Taking account of less numerical expenses in the staggered fermion, 
there is possibility that it would be numerically better than Wilson fermions,
especially as an overlap kernel \cite{PdF}.
We finally note that the analysis here does not include contribution from 
some of higher-hopping terms or higher-meson fields.
To confirm our results, we need to perform the same analysis with
these higher contributions. In the future work, we can also study detailed mass 
spectrum of mesons and possibility of small other condensation 
in the Aoki phase.

%%%%%%%%%%   ACKNOWLEDGMENTS   %%%%%%%%%%

\begin{acknowledgments}
TM is thankful to D.~Adams, M.~Creutz, M.~Golterman, T.~Izubuchi and S.~Sharpe
for the fruitful discussions.
We are thankful to P.~de~Forcrand  
for the fruitful discussions.
TK and TN are supported by Grants-in-Aid for the Japan Society 
for Promotion of Science (JSPS) Research Fellows
(Nos.
22-3314, %(T.Z. Nakano)
23-593.%, %(T. Kimura)
).
TM is supported by Grant-in-Aid for the Japan Society for Promotion of Science (JSPS) Postdoctoral
Fellows for Reseach Abroad (24-8).
This work is suppported in part by the Grants-in-Aid for Scientific Research from
JSPS (Nos. 
09J01226, %(T. Misumi)
10J03314, %(T.Z. Nakano)
11J00593, %(T. Kimrua)
23340067, %(T. Kunihiro (incl. A.Ohnishi))
24340054, %(A. Nakamura (incl. A.Ohnishi))
and
24540271. %(A. Ohnishi, K. Morita, T. Kunihiro)
), 
and by the Grant-in-Aid for the global COE program ``The Next Generation
of Physics, Spun from Universality and Emergence" from MEXT.
This work is based on fruitful discussions in the YIPQS-HPCI workshop ``New-Type of 
Fermions on the Lattice", Feb. 9-24, 2012 in Yukawa Institute for Theoretical Physics.
The authors are grateful to the organizers for giving them chances to have interest
in the present topics.
\end{acknowledgments}

\appendix

\section{spin and flavor separation}
\label{SFS}
From one-staggered field we define 16 species fields in the momentum space as
$\phi(p)_{A}\equiv\chi(p+\pi_{A})$ $(-\pi/2\leq p_{\mu} <\pi/2)$ where 
$\pi_{A}$ ($A=1,2,...,16$) being 4-dim vectors whose components take $0$ or $\pi$.
For convenience, we here consider a 16-multiplet field as $\phi(p)=(\phi(p)_{1}, 
\phi(p)_{2},\cdots, \phi(p)_{16})^{T}$.
As this 16-multiplet field has both the spinor (space-time) and the flavor (taste) indices,
we can construct two sets of clifford generators 
$\Gamma_{\mu}$ and  $\Xi_{\mu}$, which operate on spinor and flavor spaces 
in the momentum field $\phi(p)$. They satisfy the clifford algebra as 
\begin{align}
\{ \Gamma_{\mu},\Gamma_{\nu} \}=2\delta_{\mu\nu}
\ ,
\\
\{ \Xi_{\mu},\Xi_{\nu} \}=2\delta_{\mu\nu}
\ ,
\\
\{ \Gamma_{\mu}, \Xi_{\nu} \}=0
\ .
\end{align}
By using these definitions, the Dirac operator for the staggered fermion is given by 
$D_{st}=i\Gamma_{\mu}\sin p_{\mu}$ for the 16 multiplet $\phi(p)$
\footnote{The origin of the discrepancy between this form and the usual 
spin-taste representation is clearly elaborated in the reference, 
G.~P.~Lepage, [arXiv:1111.2955].}.

\section{Strong-coupling analysis for Adams-type}
\label{AdamsEff}
\begin{table*}[hbt]
\caption{Concrete forms of $\mathcal{W}_{\alpha\mu\beta\nu\gamma\rho\delta\sigma,x}^{(4)}$ in Fig. \ref{FR-A}.
}
\label{Table:U4}
\begin{tabular}{ccccc}
\hline
\hline
$\alpha$ & $\beta$ & $\gamma$ & $\delta$ & $\mathcal{W}_{\alpha\mu\beta\nu\gamma\rho\delta\sigma,x}^{(4)}$ \\ \hline
$+$ & $+$ & $+$ & $+$ & $U_{\mu,x} U_{\nu,x+\hat{\mu}} U_{\rho,x+\hat{\mu}+\hat{\nu}} U_{\sigma,x+\hat{\mu}+\hat{\nu}+\hat{\rho}}$ \\ 
$-$ & $-$ & $-$ & $-$ & $U_{\mu,x-\hat{\mu}}^\dagger U_{\nu,x-\hat{\mu}-\hat{\nu}}^\dagger U_{\rho,x-\hat{\mu}-\hat{\nu}-\hat{\rho}}^\dagger U_{\sigma,x-\hat{\mu}-\hat{\nu}-\hat{\rho}-\hat{\sigma}}^\dagger$ \\ 
$-$ & $+$ & $+$ & $+$ & $U_{\mu,x-\hat{\mu}}^\dagger U_{\nu,x-\hat{\mu}} U_{\rho,x-\hat{\mu}+\hat{\nu}} U_{\sigma,x-\hat{\mu}+\hat{\nu}+\hat{\rho}}$ \\
$+$ & $-$ & $-$ & $-$ & $U_{\mu,x} U_{\nu,x+\hat{\mu}-\hat{\nu}}^\dagger U_{\rho,x+\hat{\mu}-\hat{\nu}-\hat{\rho}}^\dagger U_{\sigma,x+\hat{\mu}-\hat{\nu}-\hat{\rho}-\hat{\sigma}}^\dagger$ \\
$+$ & $-$ & $+$ & $+$ & $U_{\mu,x} U_{\nu,x+\hat{\mu}-\hat{\nu}}^\dagger U_{\rho,x+\hat{\mu}-\hat{\nu}} U_{\sigma,x+\hat{\mu}-\hat{\nu}+\hat{\rho}}$ \\ 
$-$ & $+$ & $-$ & $-$ & $U_{\mu,x-\hat{\mu}}^\dagger U_{\nu,x-\hat{\mu}} U_{\rho,x-\hat{\mu}+\hat{\nu}-\hat{\rho}}^\dagger U_{\sigma,x-\hat{\mu}+\hat{\nu}-\hat{\rho}-\hat{\sigma}}^\dagger$ \\ 
$+$ & $+$ & $-$ & $+$ & $U_{\mu,x} U_{\nu,x+\hat{\mu}} U_{\rho,x+\hat{\mu}+\hat{\nu}-\hat{\rho}}^\dagger U_{\sigma,x+\hat{\mu}+\hat{\nu}-\hat{\rho}}$ \\ 
$-$ & $-$ & $+$ & $-$ & $U_{\mu,x-\hat{\mu}}^\dagger U_{\nu,x-\hat{\mu}-\hat{\nu}}^\dagger U_{\rho,x-\hat{\mu}-\hat{\nu}} U_{\sigma,x-\hat{\mu}-\hat{\nu}+\hat{\rho}-\hat{\sigma}}^\dagger$ \\ 
$+$ & $+$ & $+$ & $-$ & $U_{\mu,x} U_{\nu,x+\hat{\mu}} U_{\rho,x+\hat{\mu}+\hat{\nu}} U_{\sigma,x+\hat{\mu}+\hat{\nu}+\hat{\rho}-\hat{\sigma}}^\dagger$ \\ 
$-$ & $-$ & $-$ & $+$ & $U_{\mu,x-\hat{\mu}}^\dagger U_{\nu,x-\hat{\mu}-\hat{\nu}}^\dagger U_{\rho,x-\hat{\mu}-\hat{\nu}-\hat{\rho}}^\dagger U_{\sigma,x-\hat{\mu}-\hat{\nu}-\hat{\rho}}$ \\ 
$+$ & $+$ & $-$ & $-$ & $U_{\mu,x} U_{\nu,x+\hat{\mu}} U_{\rho,x+\hat{\mu}+\hat{\nu}-\hat{\rho}}^\dagger U_{\sigma,x+\hat{\mu}+\hat{\nu}-\hat{\rho}-\hat{\sigma}}^\dagger$ \\ 
$+$ & $-$ & $+$ & $-$ & $U_{\mu,x} U_{\nu,x+\hat{\mu}-\hat{\nu}}^\dagger U_{\rho,x+\hat{\mu}-\hat{\nu}} U_{\sigma,x+\hat{\mu}-\hat{\nu}+\hat{\rho}-\hat{\sigma}}^\dagger$ \\ 
$+$ & $-$ & $-$ & $+$ & $U_{\mu,x} U_{\nu,x+\hat{\mu}-\hat{\nu}}^\dagger U_{\rho,x+\hat{\mu}-\hat{\nu}-\hat{\rho}}^\dagger U_{\sigma,x+\hat{\mu}-\hat{\nu}-\hat{\rho}}$ \\ 
$-$ & $+$ & $+$ & $-$ & $U_{\mu,x-\hat{\mu}}^\dagger U_{\nu,x-\hat{\mu}} U_{\rho,x-\hat{\mu}+\hat{\nu}} U_{\sigma,x-\hat{\mu}+\hat{\nu}+\hat{\rho}-\hat{\sigma}}^\dagger$ \\ 
$-$ & $-$ & $+$ & $+$ & $U_{\mu,x-\hat{\mu}}^\dagger U_{\nu,x-\hat{\mu}-\hat{\nu}}^\dagger U_{\rho,x-\hat{\mu}-\hat{\nu}} U_{\sigma,x-\hat{\mu}-\hat{\nu}+\hat{\rho}}$ \\ 
$-$ & $+$ & $-$ & $+$ & $U_{\mu,x-\hat{\mu}}^\dagger U_{\nu,x-\hat{\mu}} U_{\rho,x-\hat{\mu}+\hat{\nu}-\hat{\rho}}^\dagger U_{\sigma,x-\hat{\mu}+\hat{\nu}-\hat{\rho}}$ \\ 
\hline
\hline
\end{tabular}
\end{table*}
In this chapter, we show the derivation of the effective potential
for the Adams-type fermion in the strong-coupling limit.
To derive the effective potential for the Adams type, 
we replace Eq.~(\ref{WZ}) by Eq.~(\ref{WZA}) in the Adams type.
\begin{align}
&\exp \left[ \sum_x N W (\Lambda) \right]
=  
\prod_x Z_x ,
\nn \\
Z_x
&= \displaystyle \int \left( \prod_{\mu \neq \nu \neq \rho \neq \sigma} \Fint \left[ U_{\mu,x}, 
U_{\mu,x+\hat{\nu}}, U_{\rho, x+\hat{\mu}+\hat{\nu}}, U_{\sigma, x+\hat{\mu}+\hat{\nu}+\hat{\rho}} \right] \right) \exp \left[ - \left( \Tr (VE^\dagger) - \Tr (V^\dagger E) \right) \right]
\ .
\label{WZA}
\end{align}
Here we represent the partition function as $4$ link integrals with 
$U_{\mu,x}$, $U_{\nu, x+\hat{\mu}}$, $U_{\rho, x+\hat{\mu}+\hat{\nu}}$, $U_{\sigma, x+\hat{\mu}+\hat{\nu}+\hat{\rho}}$. 
$V$ and $E$ in Eq.~(\ref{WZA}) 
are the matrices which include components corresponding to $1$-,
$2$-, $3$-, and $4$-link terms.
The components of $V$ and $E$ consist of link variables and fermion fields 
respectively. The concrete forms of the $V$ and $E$ for this case are given by
\begin{align}
V^{ab}_{\alpha \beta}
&= \mathrm{diag} \left(
V_1^{ab}, V_2^{ab}, \cdots, V_{11}^{ab}
\right)
\ ,
\end{align}
with%
\begin{align}
V_1
&=\mathrm{diag} \left(U_{\mu,x} \right)
\ , \\
V_2
&=\mathrm{diag} \left(
U_{\mu,x} U_{\nu,x+\hat{\mu}} U_{\rho,x+\hat{\mu}+\hat{\nu}} U_{\sigma,x+\hat{\mu}+\hat{\nu}+\hat{\rho}}
\right)
\ , \\
V_3
&=\mathrm{diag} \left(
U_{\mu,x}^\dagger U_{\nu,x} U_{\rho,x+\hat{\nu}} U_{\sigma,x+\hat{\nu}+\hat{\rho}}
\right)
\ , \\
V_4
&=\mathrm{diag} \left(
U_{\mu,x+\hat{\nu}} U_{\nu,x+\hat{\mu}}^\dagger U_{\rho,x+\hat{\mu}} U_{\sigma,x+\hat{\mu}+\hat{\rho}}
\right)
\ , \\
V_5
&=\mathrm{diag} \left(
U_{\mu,x+\hat{\rho}} U_{\nu,x+\hat{\mu}+\hat{\rho}} U_{\rho,x+\hat{\mu}+\hat{\nu}}^\dagger U_{\sigma,x+\hat{\mu}+\hat{\nu}}
\right)
\ , \\
V_6
&=\mathrm{diag} \left(
U_{\mu,x+\hat{\sigma}} U_{\nu,x+\hat{\mu}+\hat{\sigma}} U_{\rho,x+\hat{\mu}+\hat{\nu}+\hat{\sigma}} U_{\sigma,x+\hat{\mu}+\hat{\nu}+\hat{\rho}}^\dagger
\right)
\ , \\
V_7
&=\mathrm{diag} \left(
U_{\mu,x+\hat{\rho}+\hat{\sigma}} U_{\nu,x+\hat{\mu}+\hat{\rho}+\hat{\sigma}} U_{\rho,x+\hat{\mu}+\hat{\nu}+\hat{\sigma}}^\dagger U_{\sigma,x+\hat{\mu}+\hat{\nu}}^\dagger
\right)
\ , \\
V_8
&=\mathrm{diag} \left(
U_{\mu,x+\hat{\nu}+\hat{\sigma}} U_{\nu,x+\hat{\mu}+\hat{\sigma}}^\dagger U_{\rho,x+\hat{\mu}+\hat{\sigma}} U_{\sigma,x+\hat{\mu}+\hat{\rho}}^\dagger
\right)
\ , \\
V_9
&=\mathrm{diag} \left(
U_{\mu,x+\hat{\nu}+\hat{\rho}} U_{\nu,x+\hat{\mu}+\hat{\rho}}^\dagger U_{\rho,x+\hat{\mu}}^\dagger U_{\sigma,x+\hat{\mu}}
\right)
\ , \\
V_{10}
&=\mathrm{diag} \left(
U_{\mu,x+\hat{\nu}}^\dagger U_{\nu,x}^\dagger U_{\rho,x} U_{\sigma,x+\hat{\rho}}
\right)
\ , \\
V_{11}
&=\mathrm{diag} \left(
U_{\mu,x+\hat{\rho}}^\dagger U_{\nu,x+\hat{\rho}} U_{\rho,x+\hat{\nu}}^\dagger U_{\sigma,x+\hat{\nu}}
\right)
\ ,
\end{align}
\begin{align}
E^{ab}_{\alpha \beta}
&= \mathrm{diag} \left(
E_1^{ab}, E_2^{ab}, \cdots, E_{11}^{ab}
\right)
\ ,
\end{align}
and
\begin{align}
E_1
&= \mathrm{diag} \left(
D_{1,\mu}
\right)
\ , \\
E_i
&= \mathrm{diag} \left(
D_{i,\mu\nu\rho\sigma}
\right)
, \ (i=2,3,\cdots,11)
\ ,
\end{align}
where we define the operator $D$ as the fermion bilinears,
\begin{align}
\left( D_{1,\mu}^\dagger \right)^{ab} &= 
\displaystyle \frac {1}{2} \eta_{\mu,x} \chibar_x^a \chi_{x+\hat{\mu}}^b 
\ , \quad
\left( D_{1,\mu} \right)^{ab} =
 \displaystyle \frac {1}{2} \eta_{\mu,x} \chibar_{x+\hat{\mu}}^a \chi_x^b
\ , \\
\left( D_{2,\mu \nu \rho \sigma}^\dagger \right)^{ab} &= 
- s  
\chibar_x^a \chi_{x+\hat{\mu}+\hat{\nu}+\hat{\rho}+\hat{\sigma}}^b 
\ , \quad
\left( D_{2,\mu \nu} \right)^{ab} = 
s
\chibar_{x+\hat{\mu}+\hat{\nu}+\hat{\rho}+\hat{\sigma}}^a \chi_x^b 
\ , \\
\left( D_{3,\mu \nu \rho \sigma}^\dagger \right)^{ab} &= 
- s_\mu 
\chibar_{x+\hat{\mu}}^a \chi_{x+\hat{\nu}+\hat{\rho}+\hat{\sigma}}^b 
\ , \quad
\left( D_{3,\mu \nu} \right)^{ab} = 
s_\mu
\chibar_{x+\hat{\nu}+\hat{\rho}+\hat{\sigma}}^a \chi_{x+\hat{\mu}}^b 
\ , \\
\left( D_{4,\mu \nu \rho \sigma}^\dagger \right)^{ab} &= 
- s_\nu 
\chibar_{x+\hat{\nu}}^a \chi_{x+\hat{\mu}+\hat{\rho}+\hat{\sigma}}^b 
\ , \quad
\left( D_{4,\mu \nu} \right)^{ab} = 
s_\nu
\chibar_{x+\hat{\mu}+\hat{\rho}+\hat{\sigma}}^a \chi_{x+\hat{\nu}}^b 
\ , \\
\left( D_{5,\mu \nu \rho \sigma}^\dagger \right)^{ab} &= 
- s_\rho 
\chibar_{x+\hat{\rho}}^a \chi_{x+\hat{\mu}+\hat{\nu}+\hat{\sigma}}^b 
\ , \quad
\left( D_{5,\mu \nu} \right)^{ab} = 
s_\rho
\chibar_{x+\hat{\mu}+\hat{\nu}+\hat{\sigma}}^a \chi_{x+\hat{\rho}}^b 
\ , \\
\left( D_{6,\mu \nu \rho \sigma}^\dagger \right)^{ab} &= 
- s_\sigma 
\chibar_{x+\hat{\sigma}}^a \chi_{x+\hat{\mu}+\hat{\nu}+\hat{\rho}}^b 
\ , \quad
\left( D_{6,\mu \nu} \right)^{ab} = 
s_\sigma
\chibar_{x+\hat{\mu}+\hat{\nu}+\hat{\rho}}^a \chi_{x+\hat{\sigma}}^b 
\ , \\
\left( D_{7,\mu \nu \rho \sigma}^\dagger \right)^{ab} &= 
- s_{\rho+\sigma} 
\chibar_{x+\hat{\rho}+\hat{\sigma}}^a \chi_{x+\hat{\mu}+\hat{\nu}}^b 
\ , \quad
\left( D_{7,\mu \nu} \right)^{ab} = 
s_{\rho+\sigma}
\chibar_{x+\hat{\mu}+\hat{\nu}}^a \chi_{x+\hat{\rho}+\hat{\sigma}}^b 
\ , \\
\left( D_{8,\mu \nu \rho \sigma}^\dagger \right)^{ab} &= 
- s_{\hat{\nu}+\hat{\sigma}}
\chibar_{x+{\hat{\nu}+\hat{\sigma}}}^a \chi_{x+\hat{\mu}+\hat{\rho}}^b 
\ , \quad
\left( D_{8,\mu \nu} \right)^{ab} = 
s_{\hat{\nu}+\hat{\sigma}}
\chibar_{x+\hat{\mu}+\hat{\rho}}^a \chi_{x+{\hat{\nu}+\hat{\sigma}}}^b 
\ , \\
\left( D_{9,\mu \nu \rho \sigma}^\dagger \right)^{ab} &= 
- s_{\hat{\nu}+\hat{\rho}}
\chibar_{x+\hat{\nu}+\hat{\rho}}^a \chi_{x+\hat{\mu}+\hat{\sigma}}^b 
\ , \quad
\left( D_{9,\mu \nu} \right)^{ab} = 
s_{\hat{\nu}+\hat{\rho}}
\chibar_{x+\hat{\mu}+\hat{\sigma}}^a \chi_{x+\hat{\nu}+\hat{\rho}}^b
\ , \\
\left( D_{10,\mu \nu \rho \sigma}^\dagger \right)^{ab} &= 
- s_{\hat{\mu}+\hat{\nu}}
\chibar_{x+\hat{\mu}+\hat{\nu}}^a \chi_{x+\hat{\rho}+\hat{\sigma}}^b 
\ , \quad
\left( D_{10,\mu \nu} \right)^{ab} = 
s_{\hat{\mu}+\hat{\nu}}
\chibar_{x+\hat{\rho}+\hat{\sigma}}^a \chi_{x+\hat{\mu}+\hat{\nu}}^b 
\ , \\
\left( D_{11,\mu \nu \rho \sigma}^\dagger \right)^{ab} &= 
- s_{\hat{\mu}+\hat{\rho}}
\chibar_{x+\hat{\mu}+\hat{\rho}}^a \chi_{x+\hat{\nu}+\hat{\sigma}}^b 
\ , \quad
\left( D_{11,\mu \nu} \right)^{ab} = 
s_{\hat{\mu}+\hat{\rho}}
\chibar_{x+\hat{\nu}+\hat{\sigma}}^a \chi_{x+\hat{\mu}+\hat{\rho}}^b 
\ .
\end{align}
Note that $s=r \left(\epsilon \eta_5 \right)_x / (4! \cdot 16), 
s_\mu=r \left(\epsilon \eta_5 \right)_{x+\hat{\mu}} / (4! \cdot 16),$ 
and $s_{\mu+\nu}=r \left(\epsilon \eta_5 \right)_{x+\hat{\mu}+\hat{\nu}} / (4! \cdot 16)$.
Here $V_{1}$ and $E_{1}$ are $4\times 4$ diagonal matrices 
while $V_{i}$ and $E_{i}$ $(i=2,3,\cdots,11)$ are $24\times 24$ diagonal matrices.
Here we note the situation of the cancellation between the diagrams crossing the
different blocks is basically the same as the case for the Hoelbling type although
there is difference between the 2-link and 4-link hoppings. 
We can derive $W$ as a function of $\Lambda$ by using the Schwinger-Dyson equation
in a similar way to the Hoelbling type.
$\Lambda$ is 
\begin{align}
\Lambda_x
&=
\displaystyle \frac {1}{16}  
\biggl[ 
 \sum_\mu \mathcal{M}_x \mathcal{M}_{x+\hat{\mu}} 
 + \displaystyle \frac {1}{3} \sum_{\mu \neq \nu} \mathcal{M}_{x+\hat{\mu}} \mathcal{M}_{x+\hat{\mu}+\hat{\nu}} 
\ \nn \\
& + \displaystyle \frac {1}{6} \sum_{\mu \neq \nu \neq \rho} \mathcal{M}_{x+\hat{\mu}+\hat{\nu}} \mathcal{M}_{x+\hat{\mu}+\hat{\nu}+\hat{\rho}} 
  + \displaystyle \frac {1}{6} \sum_{\mu \neq \nu \neq \rho \neq \sigma} \mathcal{M}_{x+\hat{\mu}+\hat{\nu}+\hat{\rho}} \mathcal{M}_{x+\hat{\mu}+\hat{\nu}+\hat{\rho}+\hat{\sigma}} 
\biggr]
\ \nn \\
&-\left( \displaystyle \frac {r}{4! \cdot 16} \right)^2 
\sum_{\mu \neq \nu \neq \rho \neq \sigma} 
\left( 
2 \mathcal{M}_x \mathcal{M}_{x+\hat{\mu}+\hat{\nu}+\hat{\rho}+\hat{\sigma}} 
+4 \mathcal{M}_{x+\hat{\mu}} \mathcal{M}_{x+\hat{\nu}+\hat{\rho}+\hat{\sigma}} 
+2 \mathcal{M}_{x+\hat{\mu}+\hat{\nu}} \mathcal{M}_{x+\hat{\rho}+\hat{\sigma}} 
\right)
\ .
\end{align}

\section{Order parameter and zero eigenvalue of staggered-Wilson operator}
\label{zeroeigen}
We investigate the relation between the order parameter $\langle \chibar i \epsilon_x \tau_3 \chi \rangle$ for spontaneous symmetry breaking 
and zero eigenvalue of staggered-Wilson operator. 
In QCD, the chiral condensate $\langle \bar{\psi} \psi \rangle$ which is the order parameter for spontaneous breaking of chiral symmetry relates
to the zero eigenvalue of Dirac operator. It is called Banks-Casher relation \cite{BCrel}.
In Wilson fermion, the pion condensate $\langle \bar{\psi} i \gamma_5 \tau_3\psi \rangle$ 
which is the order parameter for spontaneous breaking of parity-flavor symmetry relates
to the zero eigenvalue of Wilson operator.
In staggered-Wilson fermion (two-flavor Hoelbling type), 
we derive the relation between $\langle \chibar i \epsilon_x \tau_3 \chi \rangle$ and zero eigenvalue as Wilson fermion.
Then we add the external field $H$ for order parameter to the Hoelbling-type staggered-Wilson action Eq.~(\ref{HoelS}),
\begin{align}
S_{H}(H)
&= \chibar \left[ D_{SW}(M) + i \epsilon_x \tau_3 H \right] \chi 
\ .
\end{align}
The order parameter $\langle \chibar i \epsilon_x \tau_3 \chi \rangle$ is represented as,
\begin{align}
\lim_{H \rightarrow 0} \langle \chibar i \epsilon_x \tau_3 \chi \rangle
&= -\lim_{H \rightarrow 0} \lim_{V \rightarrow \infty} \displaystyle \frac{1}{V} \mathrm{Tr} \left( i \epsilon_x \tau_3
 \displaystyle \frac{1}{D_{SW}+ i\epsilon_x \tau_3 H} \right)
\, \nn \\
&= -\lim_{H \rightarrow 0} \lim_{V \rightarrow \infty} \displaystyle \frac{1}{V} \mathrm{tr} \left[ i \epsilon_x \left( 
 \displaystyle \frac{1}{D_{SW}+ i\epsilon_x H} - \displaystyle \frac{1}{D_{SW}- i\epsilon_x H} \right) \right]
\, \nn \\
&= -\lim_{H \rightarrow 0} \lim_{V \rightarrow \infty} \displaystyle \frac{1}{V} \mathrm{tr} \left[ i \left( 
 \displaystyle \frac{1}{H_{SW}+ i H} - \displaystyle \frac{1}{H_{SW}- i H} \right) \right]
\, \nn \\
&= -i \lim_{H \rightarrow 0} \lim_{V \rightarrow \infty} \displaystyle \frac{1}{V} \sum_\lambda \langle \lambda \mid \left[  
 \displaystyle \frac{1}{\lambda+ i H} - \displaystyle \frac{1}{\lambda- i H} \right] \mid \lambda \rangle 
\, \nn \\
&= -i \lim_{H \rightarrow 0} \lim_{V \rightarrow \infty} \displaystyle \frac{1}{V} \int d\lambda \rho\left( \lambda \right) 
 \left[  
 \displaystyle \frac{1}{\lambda+ i H} - \displaystyle \frac{1}{\lambda- i H} \right]  
\, \nn \\
&= - \displaystyle \frac{2 \pi \rho(0)}{V}
\, \nn \\
&\equiv - \lim_{\epsilon \rightarrow 0} \lim_{H \rightarrow 0} \lim_{V \rightarrow \infty} \displaystyle \frac{2 \pi \rho(\epsilon)}{V}
\ ,
\end{align}
where $H_{SW}=\epsilon_x D_{SW}$ is the Hermitian operator.
$\mathrm{Tr}$ means the traces for flavor, color and space while 
$\mathrm{tr}$ means the traces for color and space. 
$\lambda$ and $\mid{\lambda}\rangle$ are the eigenvalues and eigenstates and 
$\rho(\lambda)=\sum_{\lambda^{\prime}} \delta( \lambda - \lambda^\prime )$
is the density of the state.
By this analysis,
we find the order parameter $\langle \chibar i \epsilon_x \tau_3 \chi \rangle$ for spontaneous breaking of parity-flavor symmetry relates to the zero eigenvalue of the staggered-Wilson operator
$H_{SW}$.
Also, we can derive this relation for Adams fermion in the same way.

%%%%%%%%%%   References   %%%%%%%%%%

\end{document}